\documentclass[amsmath,trackchanges, twocolumn]{aastex702}
\usepackage{graphicx}

\begin{document}

\title{Tracing M22's origins: Spatial and chemical constraints on its formation history}

\author[orcid=0000-0001-8415-8531]{Emanuele Dondoglio}
\affiliation{Physics Department, American University of Sharjah, P.O. Box 26666, Sharjah, UAE}
\affiliation{Istituto Nazionale di Astrofisica - Osservatorio Astronomico di Padova, Vicolo dell’Osservatorio 5, Padova, IT-35122}
\email[show]{edondoglio@aus.edu}

\correspondingauthor{Emanuele Dondoglio}

\author[orcid=0000-0001-7506-930X]{Antonino P. Milone}
\affiliation{Dipartimento di Fisica e Astronomia ``Galileo Galilei'', Univ. di Padova, Vicolo dell'Osservatorio 3, Padova, IT-35122}
\affiliation{Istituto Nazionale di Astrofisica - Osservatorio Astronomico di Padova, Vicolo dell’Osservatorio 5, Padova, IT-35122}
\email{antonino.milone@unipd.it}

\author[orcid=0000-0003-4861-6624]{Randa Asa’d}
\affiliation{Physics Department, American University of Sharjah, P.O. Box 26666, Sharjah, UAE}
\email{raasad@aus.edu}

\author[orcid=0000-0002-1276-5487]{Anna F. Marino}
\affiliation{Istituto Nazionale di Astrofisica - Osservatorio Astronomico di Padova, Vicolo dell’Osservatorio 5, Padova, IT-35122}
\email{anna.marino@inaf.it}

\author[orcid=0000-0002-2386-9142]{Alessandra Mastrobuono-Battisti}
\affiliation{Dipartimento di Fisica e Astronomia ``Galileo Galilei'', Univ. di Padova, Vicolo dell'Osservatorio 3, Padova, IT-35122}
\affiliation{Dipartimento di Tecnica e Gestione dei Sistemi Industriali, Università degli Studi di Padova, Stradella S. Nicola 3, I-36100, Italy}
\email{alessandra.mastrobuono@unipd.it}

\author[orcid=0009-0003-0121-7500]{Fabrizio Muratore}
\affiliation{Dipartimento di Fisica e Astronomia ``Galileo Galilei'', Univ. di Padova, Vicolo dell'Osservatorio 3, Padova, IT-35122}
\email{fabrizio.muratore@studenti.unipd.it}

\author[orcid=0000-0001-8538-2068]{Tuila Ziliotto}
\affiliation{NSF’s NOIRLab, 950 North Cherry Avenue, Tucson, AZ 85719}
\affiliation{Space Telescope Science Institute, 3700 San Martin Dr, Baltimore, MD 21218, USA}
\email{tuila.ziliotto@phd.unipd.it}

\author[orcid=0009-0000-9812-3110]{Emanuele Bortolan}
\affiliation{Dipartimento di Fisica e Astronomia ``Galileo Galilei'', Univ. di Padova, Vicolo dell'Osservatorio 3, Padova, IT-35122}
\email{emanuele.bortolan@phd.unipd.it}

\author[orcid=0000-0002-7690-7683]{Giacomo Cordoni}
\affiliation{Research School of Astronomy and Astrophysics, Australian National University, Canberra, ACT 2611, Australia}
\email{Giacomo.Cordoni@anu.edu.au}

\author[orcid=0000-0003-3153-1499]{Maria Vittoria Legnardi}
\affiliation{Dipartimento di Fisica e Astronomia ``Galileo Galilei'', Univ. di Padova, Vicolo dell'Osservatorio 3, Padova, IT-35122}
\email{mariavittoria.legnardi@studenti.unipd.it}

\author[orcid=0000-0003-1713-0082]{Edoardo P. Lagioia}
\affiliation{South-Western Institute for Astronomy Research, Yunnan University, Kunming 650500, People's Republic of China}
\email{elagioia@ynu.edu.cn}

\author[orcid=0000-0002-1128-098X]{Marco Tailo}
\affiliation{Dipartimento di Fisica e Astronomia “Augusto Righi”, Universitá di Bologna, Via Gobetti 93/2, 40129 Bologna, Italy}
\email{mrctailo@gmail.com}

\begin{abstract}

Globular clusters (GCs) are well known to host stellar populations characterized by light-element variations. A subset of Galactic GCs, beyond such 'canonical' populations, contains an additional group enriched in iron, s-process elements, and total C+N+O abundance (i.e., the anomalous stars).
We combine multi-facility photometry with APOGEE spectroscopy to investigate the spatial distribution, chemical properties, and formation history of the stellar populations in M22, with particular focus on its anomalous component. We trace the canonical and anomalous populations, together with their subpopulations, from the cluster center out to eight half-mass radii. The anomalous population becomes increasingly dominant in the outermost regions, whereas no significant radial gradients are detected among its subpopulations.
Our chemical analysis reveals light-element anticorrelations within both the canonical and anomalous components, although the latter are shifted toward higher C, N, and Al abundances. For the first time, we show that the Fe, s-process, and C+N+O enhancements among anomalous stars are not uniform but correlate with their light-element composition: the most chemically extreme anomalous stars are also the most Fe-, Ce-rich, and C+N+O-poor.
We identify a distinct red overdensity on the horizontal branch, likely populated by the most He-poor stars, and tentatively associate the extreme horizontal branch with the most chemically enriched anomalous population.
These observations are difficult to reconcile with M22 being a merger between two GCs. Instead, they qualitatively favor a self-enrichment scenario regulated by dilution, similar to that recently proposed for $\omega$Centauri, with their close chemical correspondence suggesting that they experienced analogous formation histories.

\end{abstract}



\section{Introduction}
\label{sec:1}

The presence of multiple stellar populations in globular clusters (GCs) represents one of the longest-standing problems in stellar astrophysics. Photometric and spectroscopic observations have demonstrated that massive Galactic GCs are not chemically homogeneous, but instead host distinct groups of stars characterized by different chemical compositions in elements involved in proton-capture nucleosynthesis, such as He, C, N, O, Na, Mg, and Al. Stars with chemical abundances similar to those of field stars at the same metallicity are commonly referred to as first-population (1P) stars, whereas stars enriched in the products of high-temperature hydrogen burning are classified as second-population (2P) stars \citep{bastian2018, gratton2019, milone2022}.

While light-element abundance variations are a nearly ubiquitous property of massive Galactic GCs, a small subset of clusters exhibits a more complex chemical pattern. These systems, classified as Type II GCs by \citet[][the remaining clusters being referred to as Type I]{milone2017}, display internal variations not only in light-elements but also in heavier species, including iron, s-process elements, and the total C+N+O abundance \citep[e.g.,][]{yong2008, marino2009, carretta2011}. The stars responsible for these abundance anomalies are commonly referred to as anomalous stars, in contrast to the bulk of 1P and 2P stars, which are collectively known as canonical stars. In photometric diagrams, anomalous stars typically define a redder red giant branch (RGB) and a fainter subgiant branch (SGB) than the canonical populations, reflecting their enhanced metallicity and C+N+O content \citep[e.g.,][]{milone2008, han2009, lee2009}.

Among Type II GCs, M22 (NGC\,6656) is one of the most extensively studied examples. Spectroscopic investigations have firmly established the presence of internal variations in iron, s-process elements, and total C+N+O abundance \citep{lehnert1991, dacosta2009, marino2009, marino2011b, alvesbrito2012, mckenzie2022}, while photometric studies have revealed the characteristic split RGB and SGB associated with its anomalous populations \citep{marino2009, piotto2012, lee2015, milone2017}. Intriguingly, the anomalous stars of M22 also display similar light-element patterns observed among canonical populations. In particular, they exhibit C--N, O--Na, and Mg--Al anticorrelations analogous to those that distinguish 1P and 2P stars \citep{marino2011b, lee2023}. This combination of Type II signatures and internal multiple-population patterns has motivated the development of several formation scenarios.

One possibility is that M22 formed via the merger of two originally distinct GCs, each hosting its own 1P and 2P populations. Such a merger could have occurred within the environment of a dwarf galaxy, later disrupted by its interaction with the Milky Way, as proposed by \citet{bekki2016} and supported by the analysis of \citet{lee2020}. An alternative scenario invokes a self-enrichment process, in which M22, and Type II GCs in general, experienced a more prolonged star-formation history than ordinary clusters, probably caused by such clusters being formed within a dwarf galaxy. In this framework, additional pollution episodes occurring after the formation of the canonical populations enriched the intracluster medium, giving rise to the anomalous stars \citep[e.g.,][]{marino2015, dantona2016}.

Understanding the origin of Type II GCs is a key step toward explaining why only a small fraction of Galactic GCs developed anomalous stellar populations, while the vast majority evolved as chemically simpler Type I clusters. Resolving this issue would provide fundamental constraints on the physical conditions that governed GC formation and early evolution. Moreover, both of the leading formation scenarios proposed for Type II GCs -- the merger of two originally distinct clusters and prolonged self-enrichment -- require these systems to have formed within the environment of a dwarf galaxy, a picture that is also supported by several observational studies \citep[e.g.,][]{bellazzini2008, olszewki2009, souza2026}. Deciphering the formation history of Type II GCs therefore has implications that extend well beyond the clusters themselves, offering a unique opportunity to investigate the formation and chemical evolution of dwarf galaxies, as well as their contribution to the hierarchical assembly of the Milky Way through accretion events.

In this work, we combine multiple photometric and spectroscopic datasets to disentangle the canonical and anomalous stellar populations of M22, together with their respective subpopulations, over nearly the entire extent of the cluster, from the core to approximately eight half-mass radii. Section~\ref{sec:2} presents the datasets employed in this study, while Section~\ref{sec:3} describes the photometric reduction of $u_{\rm SDSS}$ VLT Survey Telescope (VST) images. In Section~\ref{sec:4}, we introduce our population-tagging procedure based on the combined photometric and spectroscopic information. Section~\ref{sec:5} examines the radial distributions of the different populations, while Section~\ref{sec:6} investigates their chemical properties in detail. In Section~\ref{sec:7}, we explore the implications of the population complexity for the horizontal branch (HB) morphology of M22. Section~\ref{sec:8} discusses possible formation scenarios for the anomalous populations, and Section~\ref{sec:9} summarizes our main results and conclusions.

\section{Dataset}
\label{sec:2}

To investigate the different stellar populations hosted by M22, we combine photometric and spectroscopic information from several complementary datasets, spanning from the crowded cluster core to its outermost regions.

We used the Hubble Space Telescope (HST) photometric catalog of proper-motion-selected cluster members and chromosome map (ChM) published by \citet{milone2017}, based on ultraviolet and optical observations in the F275W, F336W, F438W, and F814W filters. This dataset provides full coverage of the innermost $\sim$1.2 arcmin of the cluster and ensures the high-precision photometry required to disentangle the different stellar populations in the crowded central regions.

We further exploited $u_{\rm SDSS}$-band images of M22 obtained with the VST (ID: 114.27TN, PI: E. Dondoglio). Owing to the VST one-square-degree field of view, these observations allow us to investigate the stellar populations out to a radial distance of about 30 arcmin. The dataset consists of five dithered exposures in the $u_{\rm SDSS}$ band. Details of the observations and data-reduction procedures are provided in Section~\ref{sec:3}.

To assess cluster membership outside the radial range covered by the HST catalogs, we incorporated astrometric and photometric information from the third Gaia data release \citep{gaia2023}. Candidate cluster members were selected using high-quality photometry\footnote{We adopted the renormalized unit weight error \citep[RUWE;][]{lindegren2018} as a diagnostic of the astrometric quality.}, proper motions, and parallaxes consistent with cluster membership, following the procedures described by \citet{cordoni2018} and \citet{jang2022}. The Gaia dataset will be combined with the $u_{\rm SDSS}$-band photometry from VST to separate canonical from anomalous populations in the outskirts of M22.

Our spectroscopic analysis is based on data from the Apache Point Observatory Galactic Evolution Experiment (APOGEE) Data Release 17 \citep[DR17;][]{abdurro2022}. We selected only stars with signal-to-noise ratios greater than 70 and excluded sources flagged with {\tt ASPCAPFLAG = STAR\_BAD}\footnote{APOGEE documentation: {\url{https://www.sdss4.org/dr17/}}.}. A detailed description of the selection criteria is provided by \citet{dondoglio2025}. For the stars that satisfy these requirements, we considered the abundance ratios [C/Fe], [N/Fe], [O/Fe], [Mg/Fe], [Al/Fe], [Si/Fe], [Ca/Fe], [Fe/H], and [Ce/Fe]. The APOGEE dataset includes M22 stars spanning projected distances from $\sim$0.25 to 26.5 arcmin from the cluster center.

\section{Reducing VST images in the $u_{\rm SDSS}$ band}
\label{sec:3}

The VST dataset consists of five dithered exposures in the $u_{\rm SDSS}$ band obtained on 18 June 2025, each with an exposure time of 27 s. The observations were acquired with the OmegaCAM imager, whose focal plane is composed of 32 CCD detectors, each containing $2048\times4100$ pixels\footnote{Following the OmegaCAM convention, the detectors are numbered from \#65 to \#96.}, with a plate scale of 0.21 arcsec pixel$^{\rm-1}$. Its 1$^{\rm o}\times$1$^{\rm o}$ field of view provides wide coverage of M22, well beyond the radial extent explored by previous ultraviolet photometry.

\begin{figure*}
\includegraphics[width=17cm, clip, trim={0cm 0cm 0cm 0cm}]{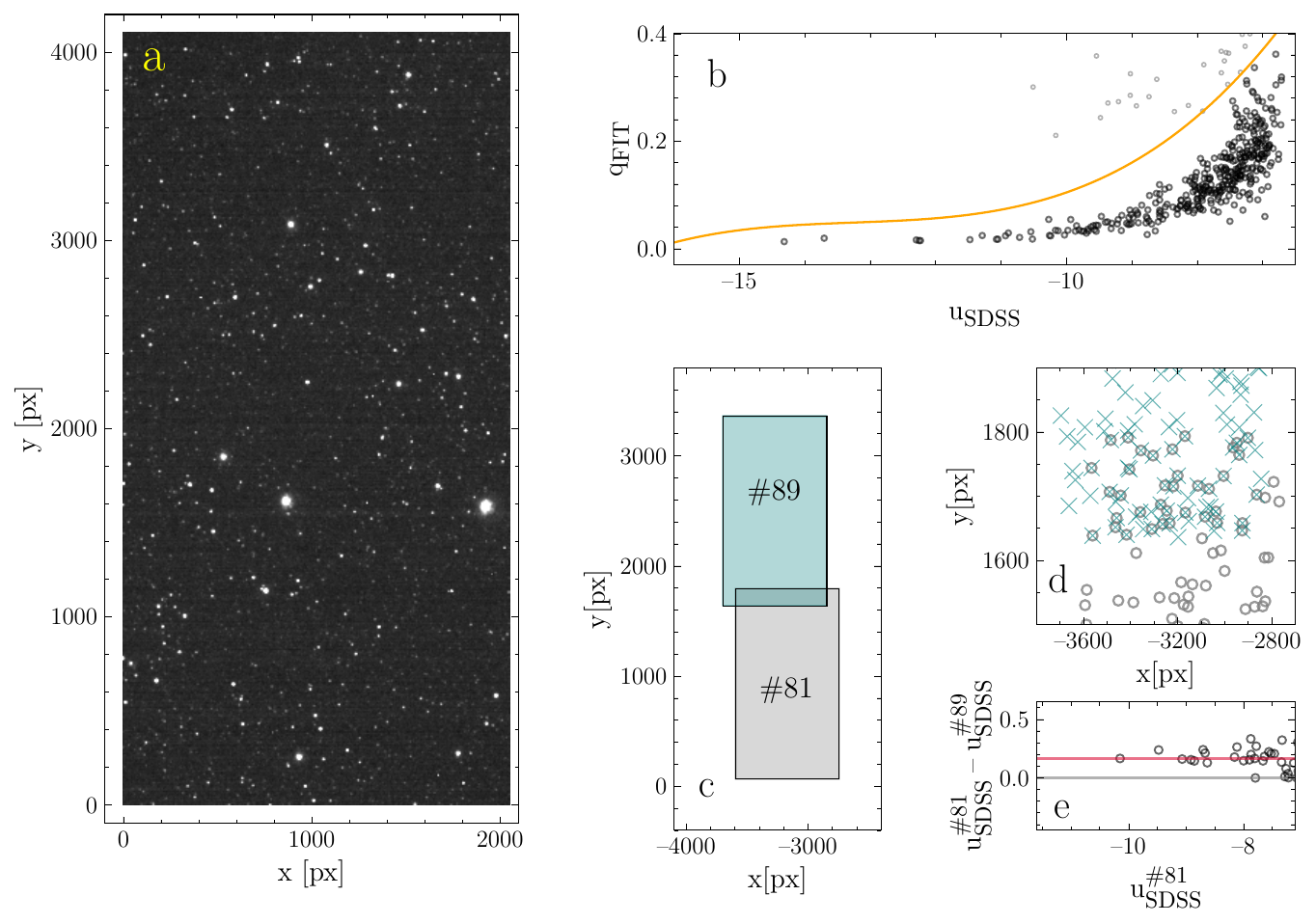}
\caption{
{\it{Panel a:}} Chip \#79 reduced OmegaCAM image of M22.
{\it{Panel b:}} $q_{\rm FIT}$ versus $u_{\rm SDSS}$ of sources measured in chip\#79, with the orange lines separating the stellar (black) from the excluded (gray).
{\it{Panel c:}} chip \#81 and \#89 footprints, colored in gray and teal, respectively.
{\it{Panel d:}} zoom-in of their coordinates (gray dots and teal crosses) in the overlapping area.
{\it{Panel e:}} chip \#81 and \#89 magnitude difference against $u_{\rm SDSS}$ of chip \#81. Gray and red lines indicate the zero-difference and the observed offset, respectively.
}
\label{fig:vst_red}
\end{figure*}

The raw images were pre-processed within the {\tt Astro-WISE} environment \citep[e.g.,][]{mcfarland2013}, following the standard reduction pipeline developed for OmegaCAM observations. Bias subtraction, bad-pixel masking, and flat-field correction were performed independently for each CCD detector. Panel~a of Figure~\ref{fig:vst_red} shows an example of one reduced image in chip \#79. 

Stellar positions and instrumental magnitudes were measured independently on each CCD using effective point-spread function (ePSF) photometry. Our reduction is based on the software developed by \citet{anderson2000} for HST imaging and subsequently adapted to ground-based observations by \citet{anderson2006}. Panel~b of Figure~\ref{fig:vst_red} illustrates the quality of the ePSF fitting through the $q_{\rm FIT}$ parameter, which quantifies how well our ePSF model reproduces the stellar profiles (values closer to zero indicate better agreement). As commonly observed in star-clusters photometry, $q_{\rm FIT}$ remains close to zero for bright stars and gradually increases toward fainter magnitudes as the signal-to-noise ratio decreases. We excluded sources with poor fits (above the orange line) from the subsequent analysis, while the retained stars are shown as black points.

The stellar catalogs derived from each detector were transformed into a common astrometric reference frame by matching their sources with Gaia photometry. The catalogs from the five dithered exposures were then cross-matched and combined following the procedure described by \citet{anderson2008}, producing a final catalog containing the average positions and magnitudes of all stars detected in multiple images.

Because each CCD has a different photometric response, detector-to-detector zero-point offsets may occur. We corrected for these differences by exploiting the dither pattern of the observations, which provides a large number of stars measured on distinct adjacent chips in different exposures. An example is shown in panel~c of Figure~\ref{fig:vst_red}, where chips \#81 and \#89 footprints from separate exposures are displayed in gray and teal, respectively, in a common pixel reference frame centered on the cluster. Their overlapping region, as illustrated in the panel~d, contains approximately 25 stars measured on both detectors. Panel~e shows the instrumental magnitude differences for these common stars as a function of magnitude. The median offset (red line) defines the relative zero-point correction between the two chips, whose value is 0.15 mag.
To place all detectors onto a homogeneous photometric system, we constructed a network of adjacent chip pairs and propagated the measured zero-point offsets throughout the entire OmegaCAM mosaic.

Instrumental magnitudes were calibrated onto the \citet{landolt1992} photometric system following the procedure described by \citet{stetson2019}, using stars from the Stetson standard-star catalog\footnote{\url{https://www.canfar.net/storage/list/STETSON/Standards}} located within the M22 field and matched to our photometric catalog.

\begin{figure}
\includegraphics[width=8.0cm, clip, trim={ 0cm 0cm 24.4cm 15cm}]{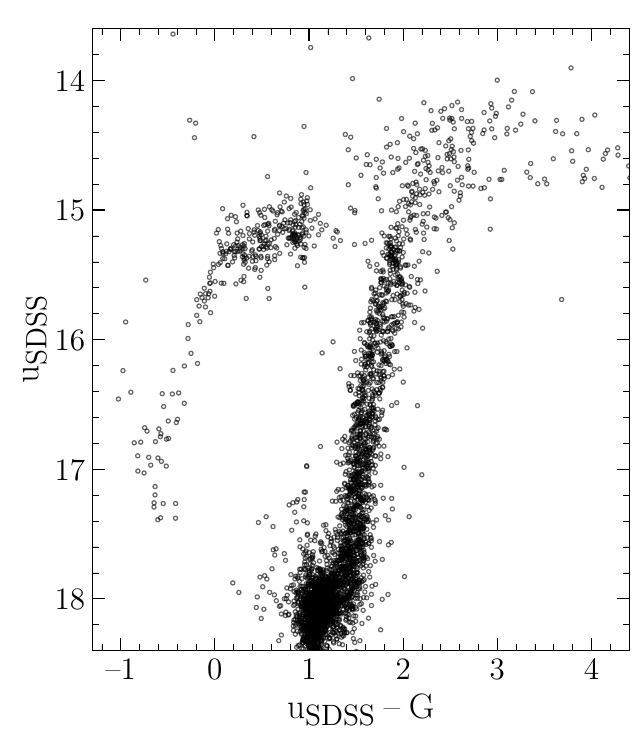}
\caption{$u_{\rm SDSS}$ versus $u_{\rm SDSS}$--$G$ CMD of M22 after proper-motion membership selection and differential-reddening correction.
}
\label{fig:vst_cmd}
\end{figure}

We then exploited Gaia proper motions to isolate M22 stars from field populations, following the procedure outlined in \citet{jang2022}. Finally, the catalog was corrected for differential reddening following the method of \citet{milone2012b}. Briefly, we used the $u_{\rm SDSS}$ versus $u_{\rm SDSS}-G$ color--magnitude diagram (CMD) to estimate, for each star, the local displacement along the reddening vector relative to a fiducial sequence defined by nearby cluster members.
The resulting corrected CMD is presented in Figure~\ref{fig:vst_cmd}. The diagram clearly reveals the characteristic split RGB of M22, the photometric signature of its canonical and anomalous stellar populations. While previous ground-based ultraviolet studies traced this feature only out to radial distances of approximately 10 arcmin, the wide field of view of the VST allows us to follow the RGB split to almost 30 arcmin from the cluster center.

\section{Following M22's populations along the RGB}
\label{sec:4}

We dedicate this Section to the coherent identification of the distinct RGB stellar populations hosted by M22, from its center out to $\sim$26.5 arcmin, by exploiting the various datasets presented in Section~\ref{sec:2}. In particular, our goal is to distinguish between canonical and anomalous stars and their respective subpopulations.

\subsection{Inside the core: HST photometry} \label{sec:4.1}

\begin{figure*}
\includegraphics[width=18cm, clip, trim={0cm 0cm 0cm 0cm}]{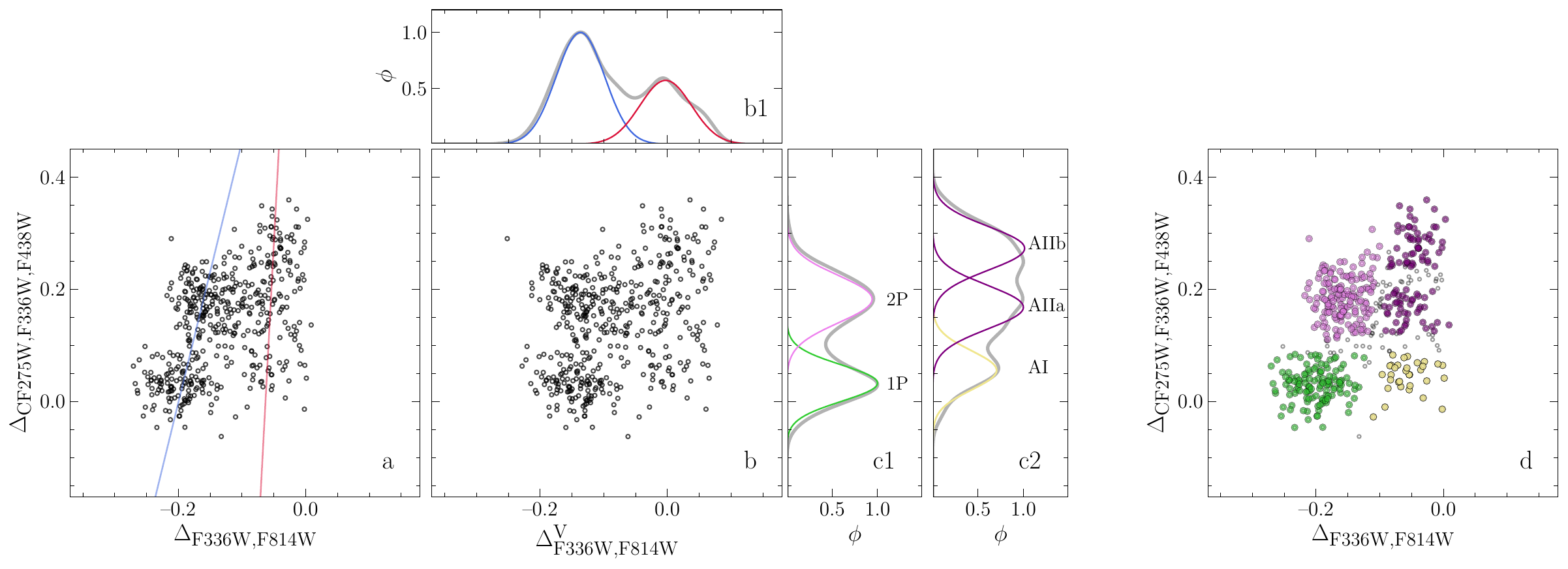}
\caption{
{\it{Panel a:}} $\Delta_{\rm C F275W,F336W,F438W}$ versus $\Delta_{\rm F336W,F814W}$ ChM of RGB stars in M22. Blue and red lines indicate best-fit lines for bonafide canonical and anomalous stars, respectively.
{\it{Panel b:}} normalized $\Delta_{\rm C F275W,F336W,F438W}$ versus $\Delta_{\rm F336W,F814W}^{\rm V}$ ChM (see text for details).
{\it{Panel b1:}} kernel density distribution of $\Delta_{\rm F336W,F814W}^{\rm V}$ (gray) line, overlapped with the two Gaussian components obtained through GMM, indicating canonical (in blue) and anomalous (in red) stars.
{\it{Panel c1:}} kernel density distribution of $\Delta_{\rm C F275W,F336W,F438W}$ for the canonical stars only, with the Gaussian components relative to 1P and 2P colored in green and violet, respectively.
{\it{Panel c2:}} same as panel c1 but for anomalous stars, where the AI and the AII (AIIa and AIIb) Gaussian distributions are colored in khaki and purple, respectively.
{\it{Panel d:}} $\Delta_{\rm C F275W,F336W,F438W}$ versus $\Delta_{\rm F336W,F814W}$ as in panel a but with stars with $>$80\% probability of belonging to 1P, 2P, AI, and AII groups colored in green, violet, khaki, and purple, respectively.
}
\label{fig:hst_chm}
\end{figure*}

In the innermost regions (within $\sim$1.2 arcmin), multiband HST photometry is available. The ChM has proven to be an effective tool for identifying chemically distinct populations in the crowded centers of GCs \citep{milone2017}. To highlight the canonical--anomalous dichotomy, we derive the ChM introduced by \citet{dondoglio2023}, which represents a variation of the classical ChM. In this formulation, the $m_{\rm F336W}$ -- $m_{\rm F814W}$ color is adopted -- in combination with the $C_{\rm F275W,F336W,F438W}$\footnote{$C_{\rm F275W,F336W,F438W} = m_{\rm F275W} - 2m_{\rm F336W} + m_{\rm F438W}$.} pseudocolor -- instead of the standard $m_{\rm F275W}$ -- $m_{\rm F814W}$, because the former is more sensitive to the chemical differences between canonical and anomalous stars. Following the procedure outlined in \citet{milone2017}, we derive the $\Delta_{\rm C F275W,F336W,F438W}$ versus $\Delta_{\rm F336W,F814W}$ ChM, shown in panel a of Figure~\ref{fig:hst_chm}.

The anomalous stars define the reddest RGB sequence, thus occupying larger x-axis coordinates (around --0.05 mag), while canonical stars populate bluer values. However, the canonical population also exhibits an internal trend, with stars at larger $\Delta_{\rm C F275W,F336W,F438W}$ displaying progressively larger $\Delta_{\rm F336W,F814W}$, broadening both populations along the x-axis.
To better highlight the canonical-anomalous dichotomy, we verticalize the x-axis of the ChM by removing the internal slope of each population. We first identify bona fide canonical and anomalous stars as those with $\Delta_{\rm F336W,F814W}$ below and above --0.10 mag, respectively, and fit the two groups with straight lines (indicated in panel~a). We then transform the ChM into a reference frame in which both best-fit lines are vertical, defining the verticalized coordinate $\Delta_{\rm F336W,F814W}^{\rm V}$. The resulting diagram is shown in panel~b, where the internal trends have been removed and both stellar groups are aligned along a common horizontal coordinate.
$\Delta_{\rm F336W,F814W}^{\rm V}$ exhibits a clear bimodality, as highlighted by the kernel density distribution shown in panel~b1. Canonical and anomalous stars produce the two peaks centered around --0.15 and 0.00 mag, respectively.

We fit the $\Delta_{\rm F336W,F814W}^{\rm V}$ distribution with a two-component Gaussian Mixture Model\footnote{We used the {\tt{sklearn.mixture}}
package \citep[][{\url{https://scikit-learn.org/stable/modules/mixture.html}}]{pedregosa2011}.} (GMM), obtaining the blue (canonical) and red (anomalous) best-fit Gaussian functions overlaid on the overall kernel density distribution in panel~b1.
We verified that the choice of two components, motivated by visual inspection, is also the one that minimizes the Akaike Information Criterion (AIC).
We measure the fraction of these two groups of stars by dividing the area subtended by each Gaussian function by the sum of the two, finding that the fraction of canonical and anomalous stars in the core are $F_{\rm CN} =$ 0.618 $\pm$0.024 and $F_{\rm AN} =$ 0.382 $\pm$0.024.

For the forthcoming subpopulations investigation, we consider as canonical and anomalous the stars with probability (provided by the best-fit GMM) of belonging to one of the two groups larger than 80\%.
Panels c1 and c2 show the kernel density distributions of $\Delta_{\rm C F275W,F336W,F438W}$ for canonical and anomalous stars (grey lines), respectively. The canonical population exhibits two well-defined peaks (at $\sim$0.04 and 0.18 mag), corresponding to the well-known 1P and 2P stellar groups. We apply the same two-components GMM procedure used above (which, again, minimizes the AIC), representing the 1P and 2P best-fit Gaussian functions in green and violet in panel c1. Their ratio, derived as for the bulk of canonical and anomalous stars, is $F_{\rm 1P} =$ 0.459 $\pm$0.029 and $F_{\rm 2P} =$ 0.541 $\pm$0.031.

Interestingly, the distribution of anomalous stars displays three distinct peaks, suggesting the presence of three chemically distinct subpopulations. Following previous studies \citep[e.g.,][]{dondoglio2023, dondoglio2026}, we define as AI the group of stars with the lowest $\Delta_{\rm C F275W,F336W,F438W}$ values, while stars at higher values are classified as AII. We further divide the AII into two subpopulations, AIIa and AIIb, corresponding to the clumps of anomalous stars centered at $\Delta_{\rm C F275W,F336W,F438W} \sim$0.17 and 0.28, respectively. The best-fit Gaussian functions are again derived with the same GMM approach, this time assuming three components (which in this case minimize the AIC). The AI Gaussian function is colored in khaki, while AIIa and AIIb are indicated in purple. The ratios of AI and AII stars are $F_{\rm AI} =$ 0.264 $\pm$0.038 and $F_{\rm AII} =$ 0.736 $\pm$0.038 ($F_{\rm AIIa} =$ 0.348 $\pm$0.038 and $F_{\rm AIIb} =$ 0.388 $\pm$0.038).

Finally, panel~d of Figure~\ref{fig:hst_chm} represents stars with probability larger than 80\% to belong to 1P, 2P, AI, and AII (AIIa+AIIb) in the $\Delta_{\rm C F275W,F336W,F438W}$ versus $\Delta_{\rm F336W,F814W}$ ChM, colored in green, violet, khaki, and purple, respectively.

\subsection{Outside the core: VST+Gaia and APOGEE} \label{sec:4.2}

\begin{figure}
\includegraphics[width=7.5cm, clip, trim={ 11cm 0cm 0cm 0cm}]{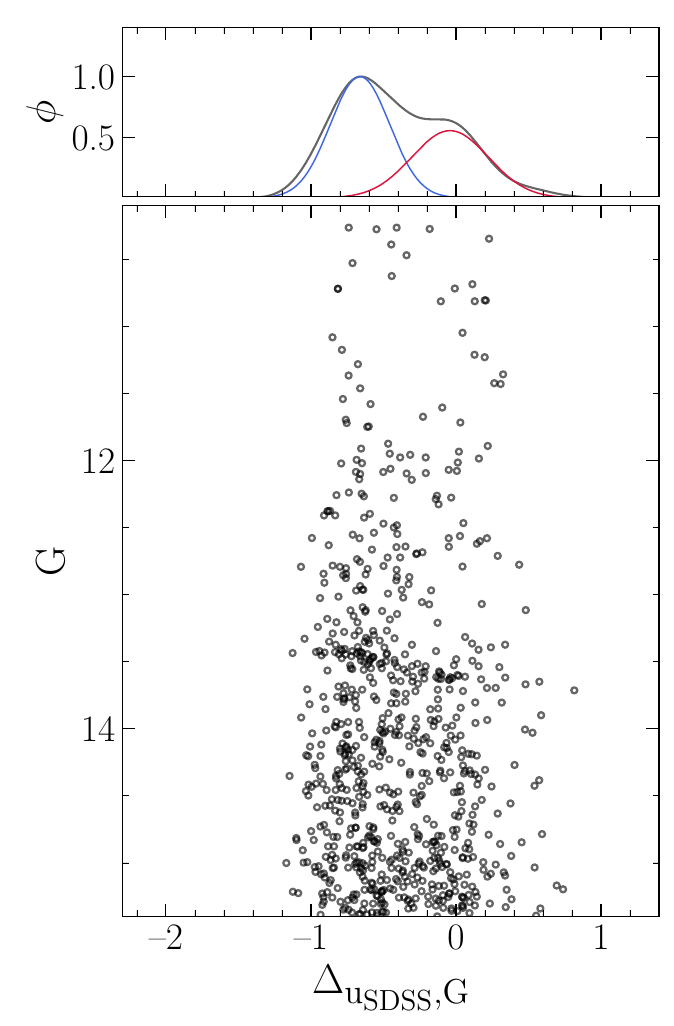}
\caption{$G$ versus the verticalized color $\Delta_{\rm u_{\rm SDSS},G}$ of RGB stars (bottom), with its kernel density distribution (top) indicated with a black line. The best-fit Gaussian representing the canonical and anomalous populations are indicated in blue and red, respectively.}
\label{fig:cmds_out}
\end{figure}

The $u_{\rm SDSS}$--$G$ color separates canonical and anomalous RGB stars outside the HST field, as shown in Figure~\ref{fig:vst_cmd}. To further enhance this separation, we followed the same procedure adopted to derive the $\Delta_{\rm F336W,F814W}$ ChM coordinate, but applied to $u_{\rm {SDSS}}$--$G$. The resulting $\Delta_{\rm u_{\rm SDSS},G}$ distribution is shown in Figure~\ref{fig:cmds_out}, where two nearly parallel sequences (i.e., canonical and anomalous stars) are located around $\Delta_{\rm u_{\rm SDSS},G}\sim$0 and $\sim$1 mag. 
As done for the HST dataset, we show the kernel density distribution (top panel) of the verticalized color, together with the best-fit Gaussian functions associated with the canonical and anomalous populations, derived through GMM fitting. With this approach, we can follow M22 populations from outside the HST coverage ($\sim$1.2 arcmin) up to $\sim$26.5 arcmin. The fraction of canonical and anomalous stars in this range, derived as in Section~\ref{sec:4.1}, are $F_{\rm CN} =$ 0.579 $\pm$0.016 and $F_{\rm AN} =$ 0.421 $\pm$0.016.

\begin{figure*}
\includegraphics[width=18.0cm, clip, trim={ 6.5cm 0cm 0cm 11cm}]{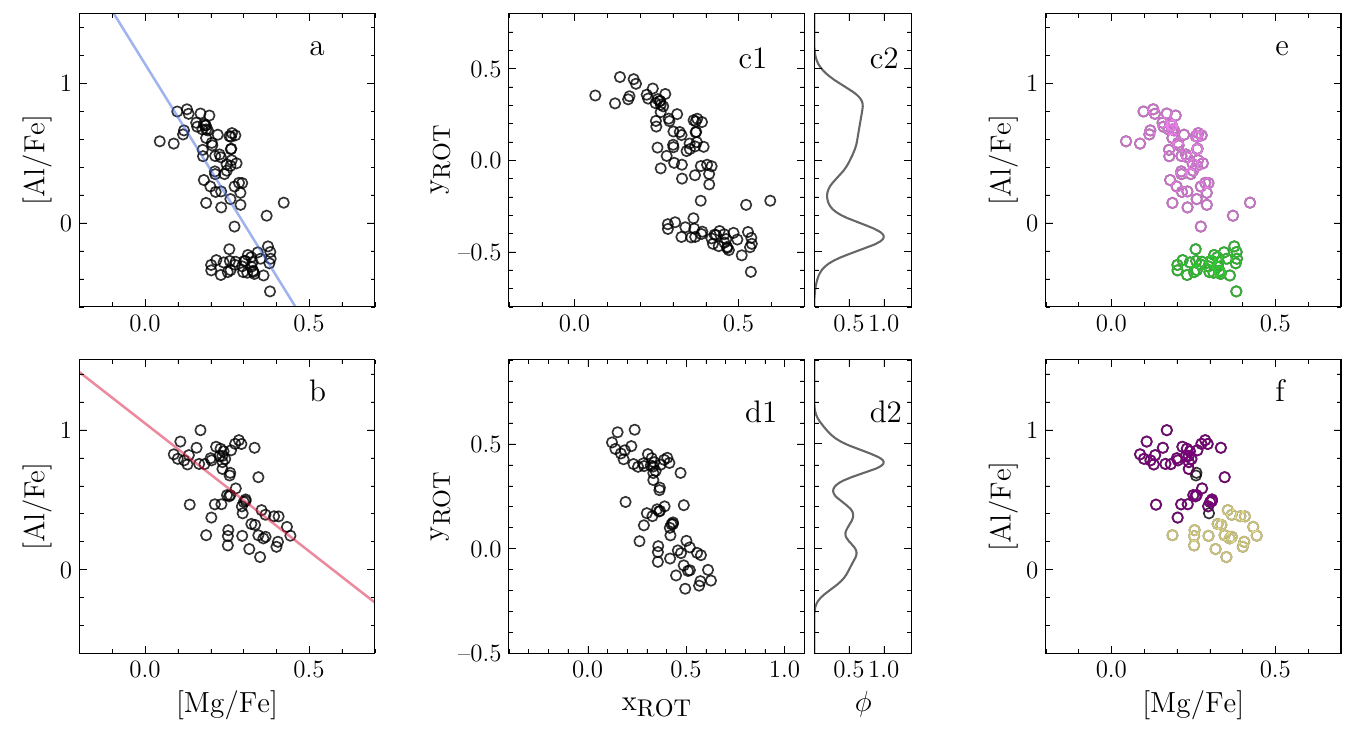}
\caption{
{\it{Panels a and b:}} [Al/Fe] versus [Mg/Fe] of RGB canonical and anomalous stars, respectively. Blue and red line indicate the best-fit straight line used to rotate their reference frame.
{\it{Panels c1 and c2:}} rotated ($x_{\rm rot}$, $y_{\rm rot}$) coordinates for canonical stars and $y_{\rm rot}$ kernel distribution.
{\it{Panels d1 and d2:}} same as c1 and c2 but for anomalous stars.
{\it{Panels e and f:}} same as panels a and b but with 1P, 2P, AI, and AII stars colored in green, violet, khaki, and purple, respectively. 
}
\label{fig:apogee}
\end{figure*}

However, $\Delta_{\rm u_{\rm SDSS},G}$ alone does not separate the 1P, 2P, AI, and AII subpopulations. To do that, we exploit the APOGEE chemical-abundance catalog, which includes a subsample of the canonical and anomalous stars tagged with VST+Gaia photometry.
Panels a and b of Figure~\ref{fig:apogee} show the [Al/Fe] versus [Mg/Fe] distributions for canonical and anomalous stars, respectively. These two groups are identified as in Figure~\ref{fig:hst_chm}, considering only stars with $>$80\% probability from the best-fit GMM to belong to one of the two populations based on their $\Delta_{\rm u_{\rm SDSS},G}$ distribution. In both cases, the diagrams reveal distinct stellar clumps, as expected from variations in light-element abundances. To identify the corresponding subpopulations, we rotate the [Al/Fe] versus [Mg/Fe] plane by the angle between the best-fit straight line to the data (blue and red lines) and the $y$-axis. The resulting rotated diagrams, expressed in the ($x_{\rm ROT}$, $y_{\rm ROT}$) reference frame, are displayed in panels b1 and c1.
Canonical stars form two clearly separated clumps, corresponding to the 1P population, centered at $y_{\rm ROT}\sim-0.4$ dex, and the 2P population, spanning approximately $y_{\rm ROT}\sim-0.2$ to 0.5 dex, as visible from the kernel density distribution shown in panel b2.
Conversely, anomalous stars exhibit three distinct peaks in their $y_{\rm ROT}$ distribution (panel c2), closely mirroring the three-populations pattern observed in Figure~\ref{fig:hst_chm}. Since AI, AIIa, and AIIb stars occupy progressively larger $\Delta_{C_{\rm F275W,F336W,F438W}}$ values -- indicative of increasing average nitrogen abundance (see Figure~8 of \citealt{milone2018}) -- we expect a similar progression to be reflected in their Al enhancement\footnote{In Section~\ref{sec:6}, we will show that indeed AI, AIIa, and AIIb have progressively larger average [N/Fe].}. We therefore identify as AI, AIIa, and AIIb the stellar groups associated with the peaks located at $y_{\rm ROT}\sim-0.02$, 0.18, and 0.42 dex, respectively. These subpopulations are separated following the same GMM-based approach adopted for Figure~\ref{fig:hst_chm}.

Finally, panels d and e of Figure~\ref{fig:apogee} show the [Al/Fe] versus [Mg/Fe] diagrams for canonical and anomalous stars, where 1P, 2P, AI, and AII(a+b) stars are highlighted in green, violet, khaki, and purple, respectively. The overall fraction for our subpopulations are the following: $F_{\rm 1P} =$ 0.358 $\pm$0.053 and $F_{\rm 2P} =$ 0.642 $\pm$0.050 among canonical stars, and $F_{\rm AI} =$ 0.310 $\pm$0.060 and $F_{\rm AII} =$ 0.690 $\pm$0.060 ($F_{\rm AIIa} =$ 0.248 $\pm$0.060 and $F_{\rm AIIb} =$ 0.442 $\pm$0.060) among anomalous stars.

Our population tagging is also in agreement with \citet{lee2020}, who also identified five stellar populations in M22, consisting of two canonical and three anomalous\footnote{In \cite{lee2020}, canonical and anomalous are dubbed G1 and G2.} groups, closely matching our classification. The relative fraction of the anomalous subpopulations are also remarkably similar within uncertainties, with the AI, AIIa, and AIIb groups contributing comparable proportions in both studies. In the canonical populations, Lee reported approximately equal fractions of 1P and 2P stars, while our analysis yields a slightly smaller 1P contribution ($\sim$40--45\%).

\section{Radial distribution of population fraction}
\label{sec:5}

We first investigate the radial behavior of the canonical and anomalous populations, quantified through the fractions $F_{\rm CN}$ and $F_{\rm AN}$. To this end, we repeat the procedure described in Section~\ref{sec:4} to derive their relative fractions in different radial bins using both the HST and VST+Gaia datasets. The bins were defined so as to contain a comparable number of stars. Specifically, we adopted two radial bins within the innermost 1.2 arcmin covered by the HST observations and five bins spanning the 1.2--26.5 arcmin radial range sampled by the outskirts photometry. Each bin contains approximately 200--220 stars.

\begin{figure}
\includegraphics[width=8.5cm, clip, trim={ 0cm 0cm 0cm 0cm}]{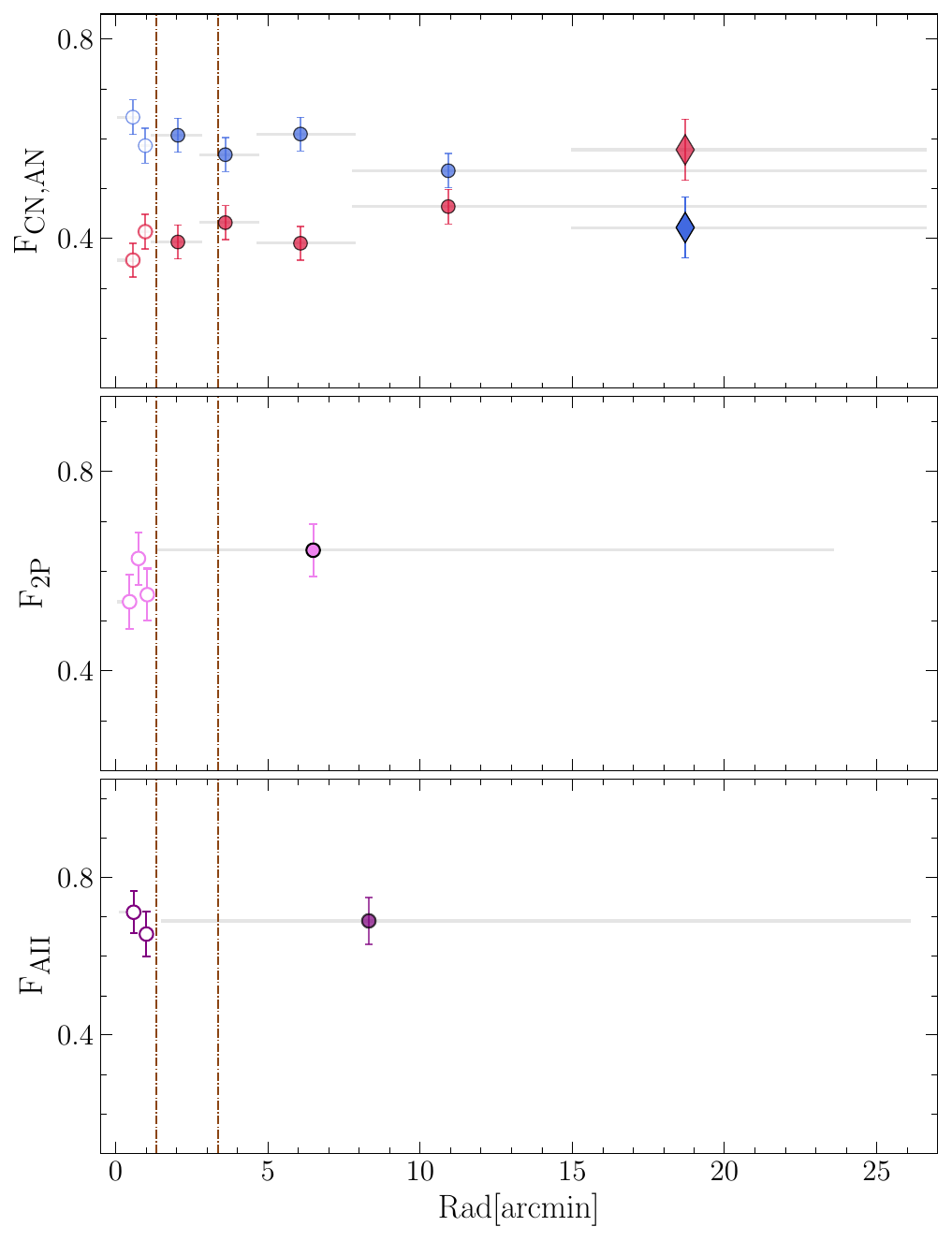}
\caption{Fraction of canonical and anomalous stars (top) in red and blue, respectively, of 2P stars compared to canonical (middle), and of AII with respect to the whole anomalous stars (bottom). Empty dots indicate measurements from HST photometry, while filled dots represent measurement with VST+Gaia (for the bulk of canonical and anomalous) or APOGEE (for the subpopulations). Brown dashed lines indicate the core and half-mass radii from \citet[][2010 edition]{harris1996}. Each measurement is associated to a horizontal gray line, indicating the reference radial range.}
\label{fig:radial}
\end{figure}

The resulting radial distributions of $F_{\rm CN}$ (blue) and $F_{\rm AN}$ (red) are shown in Figure~\ref{fig:radial}. The vertical dot-dashed brown lines indicate the core and half-mass radii \citep[from][2010 version]{harris1996}, while the horizontal gray bars associated with each point mark the radial extent covered by the corresponding bin. Open and filled symbols denote measurements from the HST and VST+Gaia datasets, respectively. Both fractions remain approximately constant out to $\sim$8 arcmin from the cluster center, where canonical and anomalous stars account for roughly 60\% and 40\%. In the outermost radial bin, however, the fractions of the two populations approach parity, with the contribution of canonical stars decreasing relative to the inner regions.
A qualitatively similar behavior was reported by \citet{lee2015}, who found the canonical component to be slightly more centrally concentrated. Interestingly, the largest deviation from a constant population ratio in their analysis occurs between $\sim$8 and 11 arcmin, where our measurements begin to depart from the nearly constant inner distribution.
To further explore this possibility in the unexplored outskirts, we repeated the analysis considering only stars located beyond 15 arcmin. Although this additional bin contains a smaller number of stars (55), and therefore is affected by larger statistical uncertainties, the resulting fractions (filled diamonds in Figure~\ref{fig:radial}) support this trend, with anomalous stars becoming the dominant population, with $F_{\rm AN} > F_{\rm CN}$ at 1-$\sigma$ level. Taken together, these results tentatively suggest that the relative distribution of canonical and anomalous stars may change in the outermost regions of M22.

We then extend the same analysis to the canonical and anomalous subpopulations by combining the tagging from Figures~\ref{fig:hst_chm} and~\ref{fig:apogee}. Outside the HST field of view, the number of stars with subpopulation identifications decreases substantially, since APOGEE abundance measurements are available for only one-fourth of the canonical and anomalous stars detected in the VST+Gaia catalog. Figure~\ref{fig:radial} shows the fraction of 2P stars relative to the total canonical population (middle panel) and the fraction of AII stars relative to the entire anomalous component (bottom panel). To maintain comparable statistics among bins, only a single radial bin outside the ChM coverage is considered.
Within the uncertainties, neither quantity exhibits significant radial variations. The limited number of APOGEE stars at large clustercentric distances prevents us from placing strong constraints on possible gradients among the individual subpopulations. In particular, the region where Figure~\ref{fig:radial} suggests a change in the relative fractions of canonical and anomalous stars is sampled by relatively few APOGEE targets. Therefore, while our data do not provide evidence for differential segregation among the 1P, 2P, AI, and AII populations, they are not sufficient to rule out the presence of such effects in the outermost regions of the cluster.

\section{The chemical composition of M22's populations}
\label{sec:6}

In this Section, we investigate in detail the chemical properties of the stellar populations identified in Section~\ref{sec:3}.

We consider ten chemical species from the APOGEE dataset, namely [C/Fe], [N/Fe], [O/Fe], [Mg/Fe], [Al/Fe], [Si/Fe], [Ca/Fe], [Fe/H], [Ce/Fe], and [(C+N+O)/Fe]. Since carbon and nitrogen abundances are known to exhibit systematic trends with stellar luminosity -- decreasing and increasing with magnitude, respectively -- we apply the procedure introduced by \citet{dondoglio2026b} to remove magnitude-dependent effects.
Briefly, we determine the 4th and 96th percentiles of the [C/Fe], [N/Fe], and [(C+N+O)/Fe] distributions in bins of $G$ magnitude, fit these fiducial boundaries with spline functions, and derive the standardized abundances [C/Fe]$_{\rm STD}$, [N/Fe]$_{\rm STD}$, and [(C+N+O)/Fe]$_{\rm STD}$ by measuring the position of each star relative to the two fiducials. This transformation removes the luminosity dependence while preserving the intrinsic abundance differences among stellar populations as below the RGB bump, allowing stars over the entire RGB to be directly compared (see Section 3.1 from Dondoglio et al. for details).

Calcium and cerium abundances display a relatively large dispersion, accompanied by substantial observational uncertainties, likely indicating the presence of residual unreliable measurements even among stars that satisfy the selection criteria described in Section~\ref{sec:2}. To mitigate this issue, we restrict the analysis to brighter stars, with higher signal-to-noise ratio hence likely more reliable measurements. Specifically, we considered [Ca/Fe] and [Ce/Fe] abundances only for stars brighter than $G$ 12.7 and 14.4 mag, respectively.

\begin{figure*}
\includegraphics[width=18cm, clip, trim={0cm 0cm 0cm 0cm}]{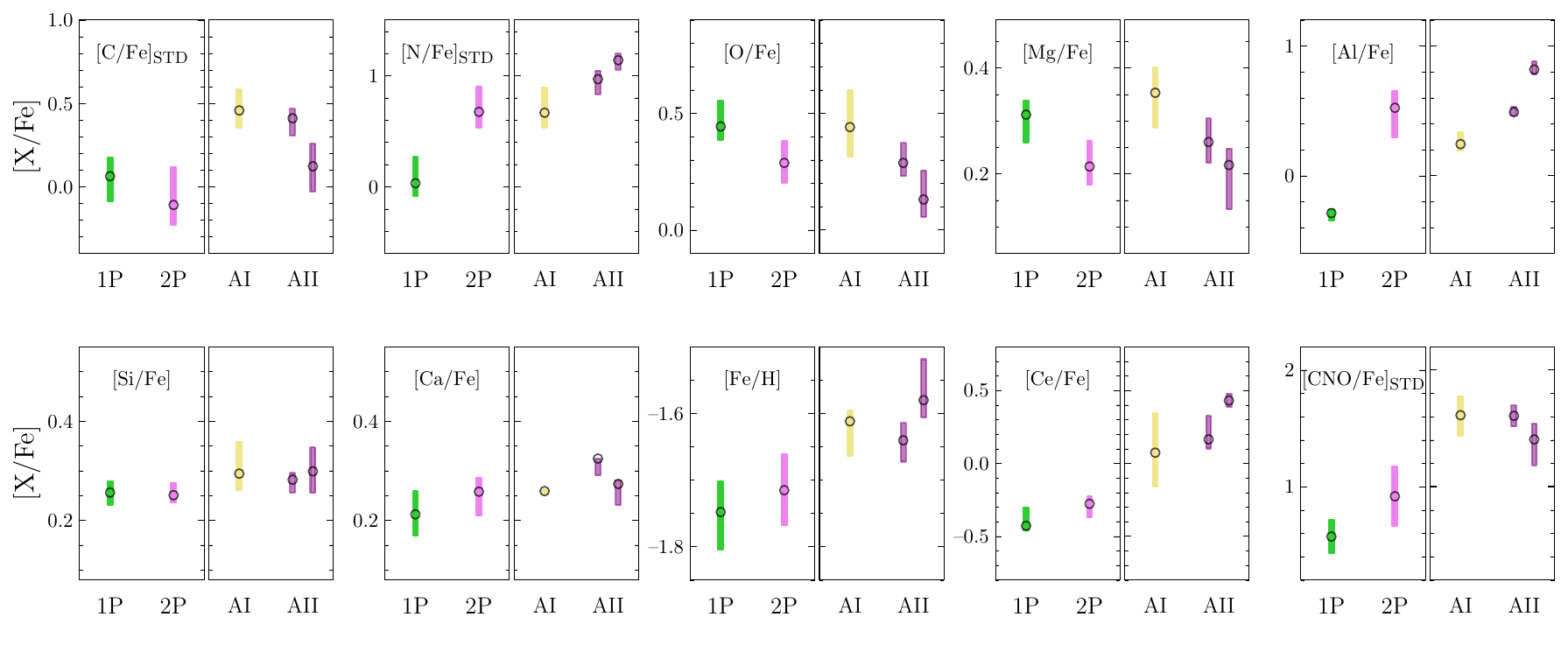}
\caption{Median abundances of ten ratios (reported in the dedicated plot). Each diagram is divided into two panels, relative to 1P and 2P (left) and AI, AIIa, and AIIb (right) stars. x-axis coordinates are shifted arbitrarily for visualization purposes. Measurements are associated with a vertical bar representing their interquartile ranges, color-coded as in Figure~\ref{fig:hst_chm}.
}
\label{fig:chemistry}
\end{figure*}

Figure~\ref{fig:chemistry} summarizes the chemical properties of the different stellar populations. For each element, the median abundance is represented by an open black dot, while the corresponding interquartile range is shown as a colored vertical bar, adopting the same color scheme introduced in Section~\ref{sec:3}. Population labels are shifted to arbitrary positions along the x-axis for visualization purposes. Each element is displayed in two panels: the left panel compares the canonical 1P and 2P populations, whereas the right panel focuses on the anomalous AI and AII populations. For the latter, the AII component is further divided into the AIIa and AIIb subgroups, represented by two separate purple symbols arranged in increasing order along the x-axis.

The upper row of Figure~\ref{fig:chemistry} presents five light-elements known to vary as part of the multiple-population phenomenon. As expected, 2P stars exhibit, on average, lower C, O, and Mg abundances, together with enhanced N and Al, relative to the 1P. The anomalous subpopulations display a qualitatively similar pattern, with AII stars being more depleted in C, O, and Mg and more enriched in N and Al than AI stars. Furthermore, AIIb stars exhibit more extreme light-element abundances than AIIa stars, lying farther from the canonical 1P composition.
The larger [N/Fe]$_{\rm STD}$ values observed in AIIb stars relative to AIIa are fully consistent with their locations along the vertical axis of the ChM, which traces nitrogen enrichment \citep{milone2018}.
The light-element abundance patterns reveal significant differences between canonical and anomalous populations. In particular, both AI and AII stars occupy systematically higher C, N, and Al abundances than the bulk of 1P+2P, while the O and Mg distributions largely overlap.

The lower row of Figure~\ref{fig:chemistry} shows the $\alpha$-elements [Si/Fe] and [Ca/Fe], both of which appear mildly enhanced in anomalous stars, with no significant differences among their subpopulations. The [Fe/H] distributions confirm the well-established enhancement of the anomalous component, whose average iron abundance is approximately 0.10 dex higher than that of canonical stars. Likewise, the s-process element [Ce/Fe] exhibits a pronounced enhancement among anomalous stars, exceeding the canonical population by approximately 0.6--0.7 dex. Finally, anomalous stars are substantially enriched in C+N+O, with [(C+N+O)/Fe]$_{\rm STD}$ values larger by roughly 1.1 dex compared to canonical stars.

Our results on the internal chemical inhomogeneities are broadly consistent with previous studies. We confirm the presence of light-element anticorrelations among both canonical and anomalous stars, with the latter following similar abundance patterns but systematically shifted toward higher C, N, and Al abundances \citep[][]{marino2011b, lee2023}. We also recover the well-established enhancements in C+N+O, iron, and s-process elements among anomalous stars \citep[e.g.,][]{marino2009, alvesbrito2012, mckenzie2022}. In particular, this work provides the first detailed characterization of the s-process enhancement in M22 based on cerium abundances. Finally, while previous investigations reported a significant calcium enhancement among anomalous stars \citep{lee2015, marino2011b}, we detect only a modest increase, likely due to the relatively small sample of stars with reliable [Ca/Fe] measurements available in our analysis.

\begin{figure*}
\includegraphics[width=18cm, clip, trim={0cm 8cm 0cm 0cm}]{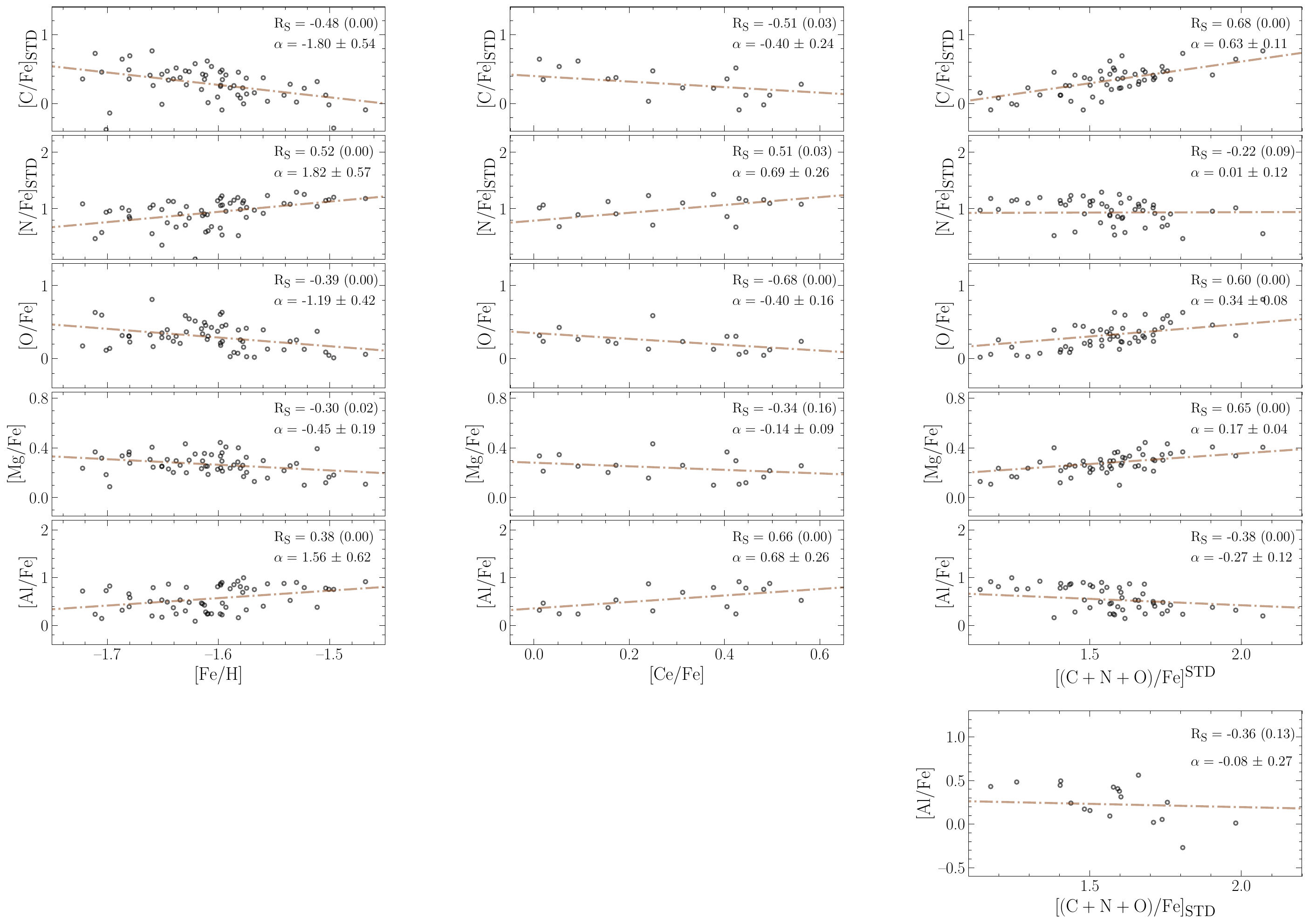}
\caption{
{\it{Left column:}} from top to bottom, [C/Fe]$_{\rm STD}$, [N/Fe]$_{\rm STD}$, [O/Fe], [Mg/Fe], and [Al/Fe] versus [Fe/H] of anomalous stars. The brown dot-dashed line in each plot indicates the best-fit straight line. 
{\it{Central and right columns:}} same as left column but with [Ce/Fe] and [(C+N+O)/Fe]$_{\rm STD}$ instead of [Fe/H], respectively. In each plot, we report the Spearman correlation coefficient (with the corresponding p-value between parenthesis) and the slope of the best-fit line with its associated uncertainty.
}
\label{fig:trend}
\end{figure*}

Intriguingly, within the anomalous component, AIIb stars exhibit marginally higher average [Fe/H] and [Ce/Fe] abundances than the other subpopulations, although their interquartile ranges partially overlap.
To investigate this behavior in greater detail, the left column of Figure~\ref{fig:trend} shows [C/Fe]$_{\rm STD}$, [N/Fe]$_{\rm STD}$, [O/Fe], [Mg/Fe], and [Al/Fe] as a function of [Fe/H] for anomalous stars only.
For each relation, we report in Figure the Spearman correlation coefficient ($R_{\rm S}$), the corresponding p-value (between parenthesis), and the slope ($\alpha$) of the best-fit line, represented with brown dot-dashed lines. The slope error has been derived via bootstrapping 1,000 times with replacements.
We find that [C/Fe]$_{\rm STD}$, [O/Fe], and [Mg/Fe] -- particularly carbon and oxygen -- decrease with increasing [Fe/H], whereas [N/Fe]$_{\rm STD}$ and [Al/Fe] increase.
Remarkably, all five elements display a fully coherent behavior: stars with the lowest C, O, and Mg abundances and the highest N and Al abundances are also those with the largest iron content, such that the most extreme anomalous stars are also the Fe-richest. Although individual relations are relatively weak (as quantified by $R_{\rm S}$), their consistency across all five light-elements suggests that the internal [Fe/H] variations are linked to the light-element abundance pattern.

The central column of Figure~\ref{fig:trend} presents the same analysis using [Ce/Fe] instead of [Fe/H]. Qualitatively similar relations emerge. Stars that are more depleted in carbon, oxygen, and magnesium and more enriched in nitrogen and aluminum tend to exhibit larger cerium abundances, mirroring the behavior observed with iron. Interestingly, similar correlations between light-element and s-process abundances have also been reported in the metal-rich GCs NGC\,6380 and Ton\,2, where [Ce/Fe] was found to correlate with [N/Fe] and, more weakly, with [Al/Fe] \citep[][]{fernandextrincado2021, fernandeztrincado2022}.

Finally, the right column compares the light-element abundances with [(C+N+O)/Fe]$_{\rm STD}$. Once again, clear trends are present (with the exception of [N/Fe]$_{\rm STD}$), but with the opposite sense relative to those involving [Fe/H] and [Ce/Fe]. In particular, [(C+N+O)/Fe]$_{\rm STD}$ increases with increasing carbon, oxygen, and magnesium abundances and decreases with increasing [Al/Fe]. Thus, stars that are more enriched in iron and cerium tend to possess lower total C+N+O abundances.

In Appendix~\ref{sec:ap1}, we repeat this analysis with the high-resolution spectroscopy from \citet{marino2011b}, showing that the same trends are independently recovered also by using a different dataset.

\section{The Horizontal Branch of M22}
\label{sec:7}

The origin of the complex HB morphologies observed in GCs remains one of the longest-standing problems in stellar astrophysics, commonly known as the second-parameter problem \citep[e.g.,][]{catelan2009}. Although a complete understanding of the distribution of HB stars in CMDs is still lacking, several observational studies indicate that 1P stars preferentially populate the reddest part of the HB, whereas 2P stars are found at bluer colors \citep[e.g.,][]{marino2011c, gratton2011}. Moreover, clusters with more extended HBs generally exhibit larger helium abundance spreads \citep{milone2018}, supporting the idea that the He-richest stars populate the hottest, bluest HB regions. Enhanced mass loss during the RGB phase is also thought to contribute by reducing the envelope mass of He-rich stars, shifting them toward bluer HB locations \citep[][]{tailo2020}.

\begin{figure}
\includegraphics[width=8.5cm, clip, trim={0cm 0cm 0cm 0cm}]{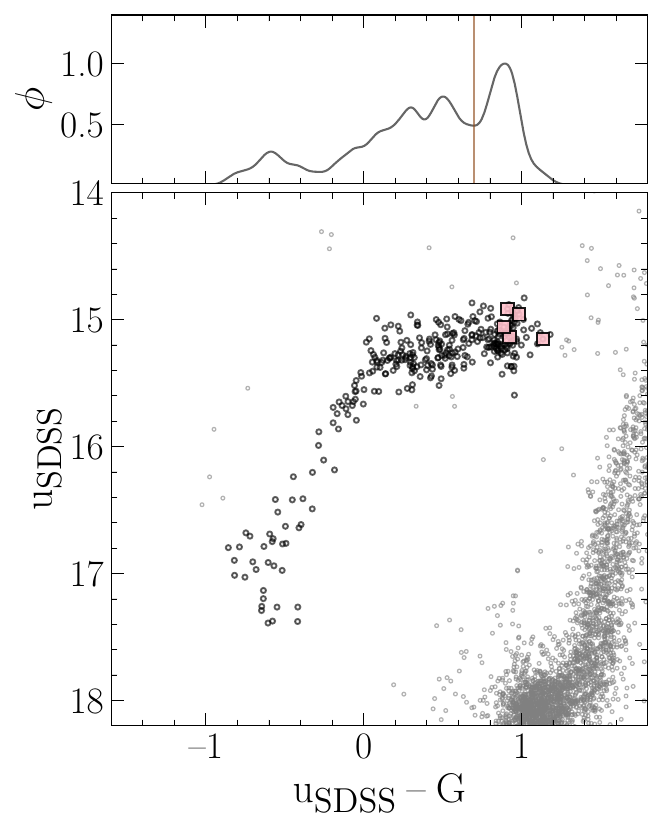}
\caption{
{\it{Bottom:}} $u_{\rm SDSS}$ versus $u_{\rm SDSS}-G$ CMD, where black points represent HB stars. Pink squares indicate stars in common with \citet{marino2013}, while the purple rectangle marks the region we expect AIIb to dominate.
{\it{Top:}} $u_{\rm SDSS}-G$ kernel density distribution of HB stars. The vertical brown line separates the reddest overdensity (see text) from the rest of the HB.
}
\label{fig:hb}
\end{figure}

In the lower panel of Figure~\ref{fig:hb}, we show the $u_{\rm SDSS}$ versus $u_{\rm SDSS}-G$ CMD centered on the HB (black dots), which exhibits a complex and extended morphology. Its reddest portion remains approximately horizontal at $u_{\rm SDSS}\simeq15.2$ mag and extends to the so-called Grundahl jump at $u_{\rm SDSS}-G\simeq0.1$ mag, a discontinuity observed among hot HB stars caused by the onset of radiative levitation in their atmospheres \citep{grundahl1999}. At bluer colors, the HB bends toward fainter magnitudes. The kernel-density distribution of $u_{\rm SDSS}-G$, shown in the top panel, reveals a prominent overdensity centered at $\sim0.9$ mag, corresponding to a distinct red group of HB stars. The brown vertical line marks the boundary adopted to separate this overdensity from the rest of the HB.

\citet{marino2013} spectroscopically analyzed six stars located on the reddest side of the HB in M22 and found chemical abundances fully consistent with those of 1P stars. Five of these objects are present in our dataset and are highlighted in the CMD as pink squares. All lie within the red overdensity, strongly suggesting that this region hosts the 1P population. However, 1P stars account for only $\sim21\%$ of the cluster population (Section~\ref{sec:4}), whereas the red HB overdensity contains $33\pm3\%$ of all HB stars. This indicates that, although the overdensity includes the 1P population, it cannot be composed exclusively of it. It is therefore likely that the most He-poor 2P and anomalous stars also contribute to this region of the HB.

The remainder of the HB is populated by progressively hotter and bluer stars belonging to the 2P and anomalous populations. Although the CMD alone does not allow a population-by-population identification, recent spectroscopic results for asymptotic giant branch (AGB) stars provide an important additional constraint. \citet{dondoglio2026b} showed that anomalous stars with [Al/Fe]$\gtrsim0.8$ dex have no AGB counterparts, implying that the AIIb population entirely avoids the AGB phase, whereas AGB descendants of the 1P, 2P, AI, and AIIa populations are all observed (their Figure~5). This behavior is characteristic of the AGB-manqué phenomenon, in which HB stars hotter than a critical effective temperature possess envelopes too thin to ascend the AGB \citep{greggio1990}. Since only the AIIb population exhibits this behavior, and the stars that skip the AGB are generally the hottest ones on the HB, it is tempting to associate the bluest and faintest HB region in M22 with these most chemically extreme anomalous stars.

\section{On the origin of anomalous stars in M22}
\label{sec:8}

The chemical patterns presented in Section~\ref{sec:6} provide a set of observational constraints that any successful formation scenario for Type II GCs must reproduce. In particular, our analysis identifies four key properties of the anomalous populations in M22: (i) anomalous stars are enriched in iron, s-process elements, and total C+N+O relative to the canonical populations; (ii) they exhibit internal light-element variations analogous to those observed between canonical 1P and 2P stars; (iii) their light-element distributions are systematically shifted toward higher carbon, nitrogen, and aluminum abundances than the canonical populations; and (iv) their iron, s-process, and C+N+O abundances are not constant, but correlate with the degree of light-element enrichment (Figure~\ref{fig:trend}). In the following, we assess whether the two main scenarios proposed for the origin of Type II GCs -- the merger of two initially distinct GCs and prolonged self-enrichment within a single stellar system -- can naturally account for these observational constraints. We emphasize that, given the present uncertainties in the theoretical modeling of Type II clusters, our discussion is necessarily qualitative.

\subsection{The merger hypothesis}
\label{sec:8.1}

In the merger scenario, the canonical and anomalous populations originate from two independent GCs formed within the same dwarf galaxy, which subsequently merged before being accreted by the Milky Way \citep[e.g.,][]{bekki2016,lee2020, bekki2026}. Within this framework, the global iron, s-process, and C+N+O enhancement of anomalous stars naturally reflects the chemical evolution of the host dwarf galaxy: the anomalous progenitor cluster would have formed later than the canonical one, from gas that had already experienced additional chemical enrichment. The presence of internal light-element variations among anomalous stars follows naturally if the anomalous component originated in an independent GC that developed its own multiple-population phenomenon before merging with the canonical cluster.

However, two observational findings are difficult to reconcile with this interpretation:

\begin{itemize}

\item In this framework, AI stars would naturally correspond to the first-population stars of the anomalous progenitor cluster. Their chemical composition is therefore expected to resemble that of ordinary 1P stars at the same metallicity. Instead, AI stars exhibit unusually high nitrogen and aluminum abundances (Figure~\ref{fig:chemistry}), much closer to those of canonical 2P stars. Even sodium -- not included in our analysis -- mirrors this behavior, as shown by \citet[][see their Figure~14]{marino2011b}. To illustrate this discrepancy, Figure~\ref{fig:ai_vs_1p} shows the median [Al/Fe] abundance of 1P stars measured in 16 Galactic GCs by \citet{dondoglio2025,dondoglio2026} using APOGEE data (error bars indicate the corresponding standard deviations). The average [Al/Fe] of 1P stars increases with [Fe/H], as expected from Galactic chemical evolution \citep[e.g.,][]{matteucci2021,kobayashi2025}, and the 1P population of M22 (green dot) follows this relation. In contrast, the AI population (khaki dot) lies about 0.5 dex above the Galactic trend. Even if M22 formed within a dwarf galaxy rather than the Milky Way, such high [Al/Fe] abundances are not observed among dwarf-galaxy field stars at comparable, or even higher, metallicities \citep[e.g.,][]{hasselquist2017,shetrone2026,xu2026}.

\begin{figure}
\includegraphics[width=8cm, clip, trim={0cm 0cm 0cm 0cm}]{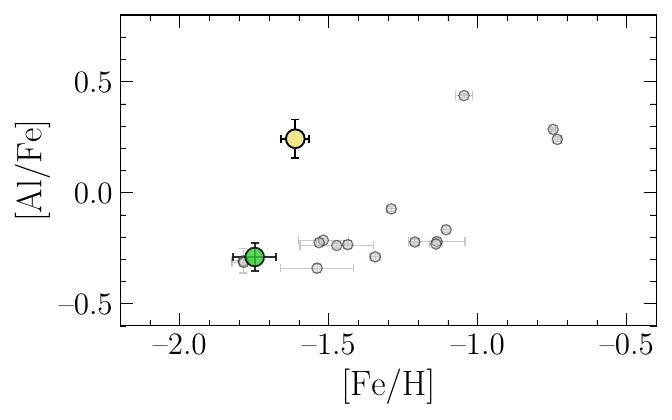}
\caption{Median [Al/Fe] versus [Fe/H] of 1P stars measured by \citet{dondoglio2025, dondoglio2026} with APOGEE for 16 Galactic GCs. The same for 1P and AI stars are indicated with the green and khaki dots, respectively.}
\label{fig:ai_vs_1p}
\end{figure}

\item A second difficulty arises from the internal chemical trends displayed in Figure~\ref{fig:trend}. The anomalous stars exhibiting the largest light-element differences relative to AI are also slightly more enriched in iron and cerium while displaying lower [(C+N+O)/Fe]. This behavior has no analogue among Type I GCs, where 2P stars span a wide range of light-element abundances while maintaining essentially constant iron, s-process, and C+N+O abundances. If the AI and AII populations originated within an independent GC analogous to present-day Type I clusters, these internal heavy-element trends would not be expected.

\end{itemize}

Overall, although the merger hypothesis naturally explains the existence of anomalous stars, their global iron-, s-process-, and C+N+O enrichment, and the presence of multiple populations within the anomalous component, it struggles to reproduce both the peculiar chemistry of the AI population and the internal correlations linking heavy and light-elements. Reconciling these observations would require the anomalous progenitor cluster to have experienced an enrichment history substantially different from that inferred for ordinary Type I Galactic GCs. While our strictly qualitative considerations cannot robustly discard this idea, our observations seems to disfavor the merging interpretation for M22.

\subsection{The self-enrichment hypothesis}
\label{sec:8.2}

An alternative possibility is that M22 formed as a single stellar system that experienced a more prolonged star-formation history than Type I GCs. In this framework, the canonical populations formed first, while anomalous stars originated later from gas additionally enriched by polluters operating on longer timescales. Candidate enrichment sources include intermediate-mass AGB stars together with Type Ia supernovae, acting after approximately 100 Myr after a GC form \citep{dantona2016}, or massive stars, possibly aided by stellar rotation, operating over only a few Myr \citep{gieles2025}. In either case, the dwarf-galaxy environment would provide the deep gravitational potential required to retain these ejecta, explaining the global iron, s-process, and C+N+O enrichment of the anomalous populations.

The remaining chemical properties can be interpreted within the dilution scenario recently proposed by \citet{dondoglio2026} for the anomalous populations of $\omega$Centauri \citep[see also][]{marino2011a}. In this picture, the most chemically extreme anomalous stars (AIIb) form first from gas dominated by the ejecta of the anomalous polluters. Subsequently, the remaining anomalous populations (AIIa and AI) originate from the same enriched reservoir progressively diluted with intracluster gas having approximately 1P-like composition. Notably, a similar dilution process is already thought to regulate the formation of canonical 2P stars in virtually all proposed scenarios for ordinary GCs \citep[e.g.,][]{dantona2016,renzini2023,gieles2025}.

This framework is consistent with anomalous stars exhibiting internal light-element variations: just as different dilution factors generate the canonical 1P--2P anticorrelations, they produce a similar distribution among AI and AII, thus explaining the internal light-element inhomogeneities. An attractive feature of this scenario is that dilution simultaneously affects both light and heavy elements. Consequently, stars formed from less diluted ejecta are expected to be more enriched not only in aluminum and nitrogen, but also in iron and s-process elements. This idea qualitatively explains the observed increase of [Fe/H] and [Ce/Fe] toward the most chemically extreme anomalous stars. Furthermore, if AI stars would still retain a substantial fraction of enriched material (as suggested by them being Fe-richer than 1P), their systematically higher nitrogen and aluminum abundances relative to the canonical populations arise naturally.

\subsubsection{On the composition of the diluting gas} \label{sec:8.2.2}

Carbon, however, poses a significant challenge. While AIIb stars exhibit [C/Fe]$_{\rm STD}$ values comparable to those of 1P stars, both AIIa and AI are substantially more carbon-rich than the canonical population. Such a behavior cannot be reproduced by a simple mixture of AIIb- and 1P-like material. Moreover, if AI and AIIa formed through increasing dilution of the gas that produced AIIb, one would expect the total C+N+O abundance to decrease progressively toward the canonical value, as observed for [Fe/H] and [Ce/Fe]. Instead, the right column of Figure~\ref{fig:trend} reveals the opposite behavior.

One possible way to reconcile these observations is to invoke an additional source of carbon enrichment contributing to the diluting medium. Intriguing candidates are low-mass AGB stars (initial masses of $\sim$1--3\,M$_{\odot}$), which efficiently produce carbon without substantially modifying the other light-elements \citep{ventura2022} and begin releasing their ejecta shortly after the epoch dominated by intermediate-mass AGB stars and Type Ia supernovae. In this picture, the diluting gas would progressively incorporate an increasing fraction of C-rich ejecta as the more Al-poor anomalous populations formed. Such a mechanism may explain the increase in [C/Fe]$_{\rm STD}$ from AIIb to AI and may also account for the opposite behavior of [(C+N+O)/Fe]$_{\rm STD}$. Additional observational support comes from the Mg isotopic ratios measured by \citet{mckenzie2024}, which indicate that ejecta from AGB stars in the $\sim$1--3\,M$_{\odot}$ range contributed to the material from which the M22 anomalous stars formed.

A potential difficulty, however, is that low-mass AGB stars are also expected to synthesize significant amounts of s-process elements \citep[e.g.,][]{cristallo2009}, whereas our analysis hints at a mild decrease in [Ce/Fe] from the most chemically extreme AIIb stars toward AI. Dedicated nucleosynthetic modeling will therefore be required to assess whether the combined contributions of low- and intermediate-mass AGB stars together with Type Ia supernovae can simultaneously reproduce the complete set of observational constraints identified in this work.

Alternatively, if the anomalous enrichment was instead driven by massive stars, the available timescale for star formation would be considerably shorter, of the order of only a few Myr. In that case, explaining the observed carbon and C+N+O enhancements would require a class of massive-star polluters capable of producing substantial carbon enrichment without simultaneously generating significant s-process enrichment or incompatible light-element abundances. Carbon yields from massive stars remain particularly uncertain \citep[see the discussion in][]{Asad24}, and dedicated calculations tailored to the chemical patterns of M22 will be necessary to evaluate whether this alternative scenario is viable.

\subsubsection{A comparison with $\omega$Centauri}
\label{sec:8.2.2}

The dilution scenario discussed above was originally proposed to explain the anomalous populations of $\omega$Centauri. An important question is therefore whether the same framework can simultaneously explain the similarities and differences between both clusters.

Several works in the literature compared $\omega$Centauri and M22, suggesting that they may share a similar formation history \citep[e.g.,][]{norris1983, lee2009, dacosta2011}. In this Section, we do that by combining our data with the APOGEE catalog from \citet{dondoglio2026}. These two GCs are suitable for a direct comparison because their 1P stars have nearly identical chemical compositions (i.e., they formed from similar natal gas), thus minimizing the impact of metallicity-dependent nucleosynthetic yields and allowing a direct confrontation of their multiple-population patterns.

\begin{figure*}
\includegraphics[width=18cm, clip, trim={0cm 0cm 0cm 0cm}]{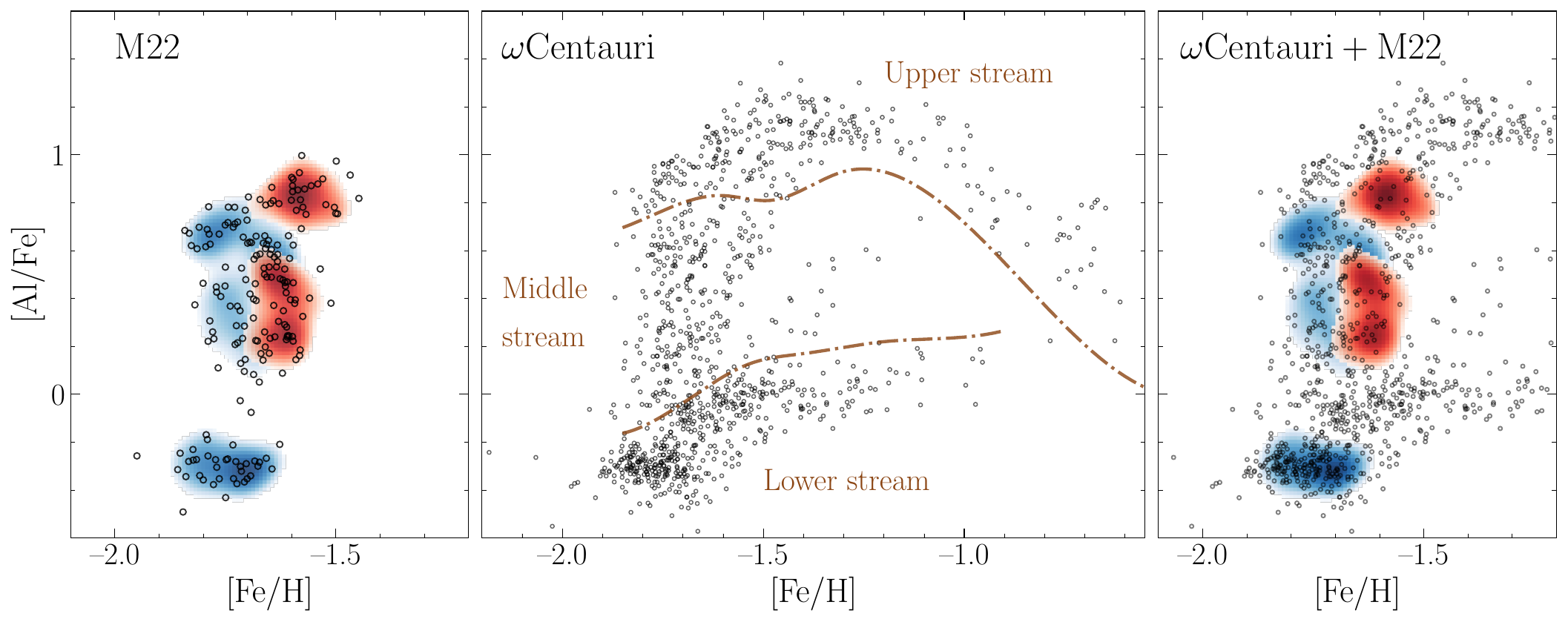}
\caption{
{\it{Left:}} [Al/Fe] versus [Fe/H] of M22. Blue and red colormaps indicate the soothed distribution of its canonical and anomalous stars, respectively.
{\it{Middle:}} [Al/Fe] versus [Fe/H] of $\omega$Centauri. Brown dash-dotted lines separate the lower, middle, and upper streams.
{\it{Right:}} same but zoomed-in in $\omega$Centauri Fe-rich end, overlayed to the canonical and anomalous stars smoothed distributions.}
\label{fig:wcen}
\end{figure*}

The left panel of Figure~\ref{fig:wcen} displays the [Al/Fe] versus [Fe/H] of M22 RGB stars, in which we overlay the smoothed distribution of its canonical and anomalous stars with blue and red colormaps, respectively. The middle panel illustrates the same plot for the RGB stars of $\omega$Centauri, where we separated with brown lines the three main streams \citep[as defined by][]{marino2019, dondoglio2026}: (i) the lower stream, composed of 1P and the most Al-poor tail of anomalous stars, (ii) the upper stream, which comprises the most Al-rich (2P and anomalous) stars along the whole [Fe/H] range, and (iii) the middle stream, made by Al-intermediate 2P and anomalous stars. In the right panel we overlay $\omega$Centauri abundances with the smoothed distributions of canonical and anomalous stars in M22 (as derived in the left panel).

M22 appears as a 'subset' of $\omega$Centauri, with the 1P of both clusters overlapped at ([Fe/H], [Al/Fe])$\sim$(--1.75, --0.35). 2P and anomalous stars in M22 broadly follow the distribution of the middle stream of $\omega$Centauri up to [Fe/H]$\sim$--1.55 dex, with the most Al-rich anomalous stars being slightly Fe-richer. The two main differences are the lack of lower (beside the 1P) and upper stream stars.
In this picture, the absence of the anomalous lower stream may be a consequence of the lower initial mass of M22, whose shallower gravitational potential was probably unable to retain the core-collapse supernova ejecta thought to produce this population in $\omega$Centauri \citep{marino2011a,mason2025}. 

Interestingly, the lack of upper-stream stars may be a consequence of dilution. Indeed, more massive GCs host more chemically extreme 2P stars at fixed metallicity \citep[e.g.,][]{carretta2010, dondoglio2025}, with theoretical models predicting weaker dilution in more massive clusters \citep[e.g., Figure~12 of][]{gieles2025}. In this picture, dilution would have been more efficient in the least-massive M22, thus not allowing the production of the poorly-diluted upper-stream stars. The fact that both canonical and anomalous stars follow this same behavior favors the idea that they were shaped by the same underlying enrichment+dilution mechanism.

Overall, the remarkable chemical similarity between M22 and $\omega$Centauri strengthens the possibility that the two Type II GCs experienced qualitatively similar self-enrichment histories. Their present-day differences can be understood primarily as the consequence of their different initial masses, rather than requiring fundamentally different formation mechanisms.

\section{Summary and Conclusions}
\label{sec:9}

In this study, we combined wide-field photometric datasets (HST, VST, and Gaia), covering M22 from its central regions to nearly eight times its half-mass radius, with APOGEE spectroscopy to investigate the spatial distribution, chemical properties, and possible origin of its stellar populations. Our main results can be summarized as follows.

\begin{itemize}

    \item Using photometric diagnostics specifically designed to separate the canonical and anomalous populations, we traced the two groups from the cluster center to nearly 30 arcmin. By combining these data with APOGEE abundances, we further identified the canonical subpopulations (1P and 2P) and three distinct anomalous groups, which we designate AI, AIIa, and AIIb.

    \item The radial distributions of canonical and anomalous stars remain indistinguishable up to approximately two half-mass radii. The fraction of canonical stars decreases only in the outermost regions, corresponding to a relative increase in the contribution of anomalous stars. No significant radial gradients are detected among the individual subpopulations.

    \item We derived the chemical composition of all five stellar populations identified in M22. The canonical populations exhibit the well-known light-element abundance patterns associated with the multiple-population phenomenon, with 2P stars being depleted in C, O, and Mg and enriched in N and Al relative to 1P stars. The anomalous populations display analogous internal variations, although shifted toward systematically larger C, N, and Al abundances than the canonical stars. In addition to confirming the established enhancements in iron, s-process elements, and total C+N+O among anomalous stars, we show for the first time that these quantities correlate with the light-element abundances. In particular, the most chemically extreme anomalous stars are also the most Fe- and Ce-rich, while exhibiting the lowest [(C+N+O)/Fe] values.

    \item The HB of M22 exhibits a complex morphology in the $G$ versus $u_{\rm SDSS}-G$ CMD, reflecting the chemical diversity observed along the RGB. We identify a distinct red overdensity likely populated by 1P stars, together with the least chemically extreme 2P (and possibly AI) stars. Moreover, based on the different chemical patterns observed between RGB and AGB stars, we argue that the extreme, hottest HB component is dominated by AIIb stars, which represent the most chemically enriched anomalous population.

    \item The chemical properties of the anomalous populations present significant challenges to a merger scenario. In particular, the unusual composition of AI stars and the internal correlations involving Fe, Ce, C+N+O, and light-elements are difficult to reconcile with the interpretation of AI and AII as the counterparts of 1P and 2P stars in a second merged cluster. Conversely, the observed abundance patterns show qualitative similarities with those recently identified in $\omega$Centauri and appear more naturally explained within a self-enrichment framework regulated by dilution.

\end{itemize}

Future works, involving other Type II Gcs and tailored nucleosynthetic modeling, will be crucial to robustly test the proposed scenario, constraining the the origin of the polluter(s), address a possible mass-budget problem for such self-pollution process \citep[as recently raised by][]{bekki2026}, and account for the cluster-to-cluster variability observed within this class of GCs.

\noindent author year, title, version, publisher, prefix:identifier\\

\begin{acknowledgments}

This work has been funded by the European Union – NextGenerationEU RRF M4C2 1.1 (PRIN 2022 2022MMEB9W: “Understanding the formation of globular clusters with their multiple stellar generations”, CUP C53D23001200006).
This paper represents the opinions of the authors and does not mean to represent the position or opinions of the American University of Sharjah. E. P. L. acknowledges support by Special Project for High-End Foreign Experts "Xingdian" Funding from Yunnan Province and National Key R\&D Program of China Grant (No. 2024YFA1611601).
Based on data collected with the INAF VST telescope at the ESO Paranal Observatory

\end{acknowledgments}

%
\facilities{HST(WFC3 and ACS), VST(OmegaCAM), Apache Point Observatory, Gaia.}


\appendix

\section{Internal trends in anomalous stars from Marino et al. (2011)} \label{sec:ap1}

To assess the robustness of the internal chemical trends identified in Figure~\ref{fig:trend}, we repeated the same analysis using the high-resolution spectroscopic abundances published by \citet{marino2011b}. Although this sample comprises fewer stars than the APOGEE dataset, it is based on spectra with substantially higher resolving power ($R\sim38,000$--60,000, compared to APOGEE's $R\sim22,500$), thus providing an important independent test of our results. In addition, the two datasets probe complementary spectral regions: APOGEE derives abundances from near-infrared $H$-band spectra, whereas \citet{marino2011b} employed optical spectroscopy.

\begin{figure*}
\includegraphics[width=18cm, clip, trim={0cm 0cm 0cm 0cm}]{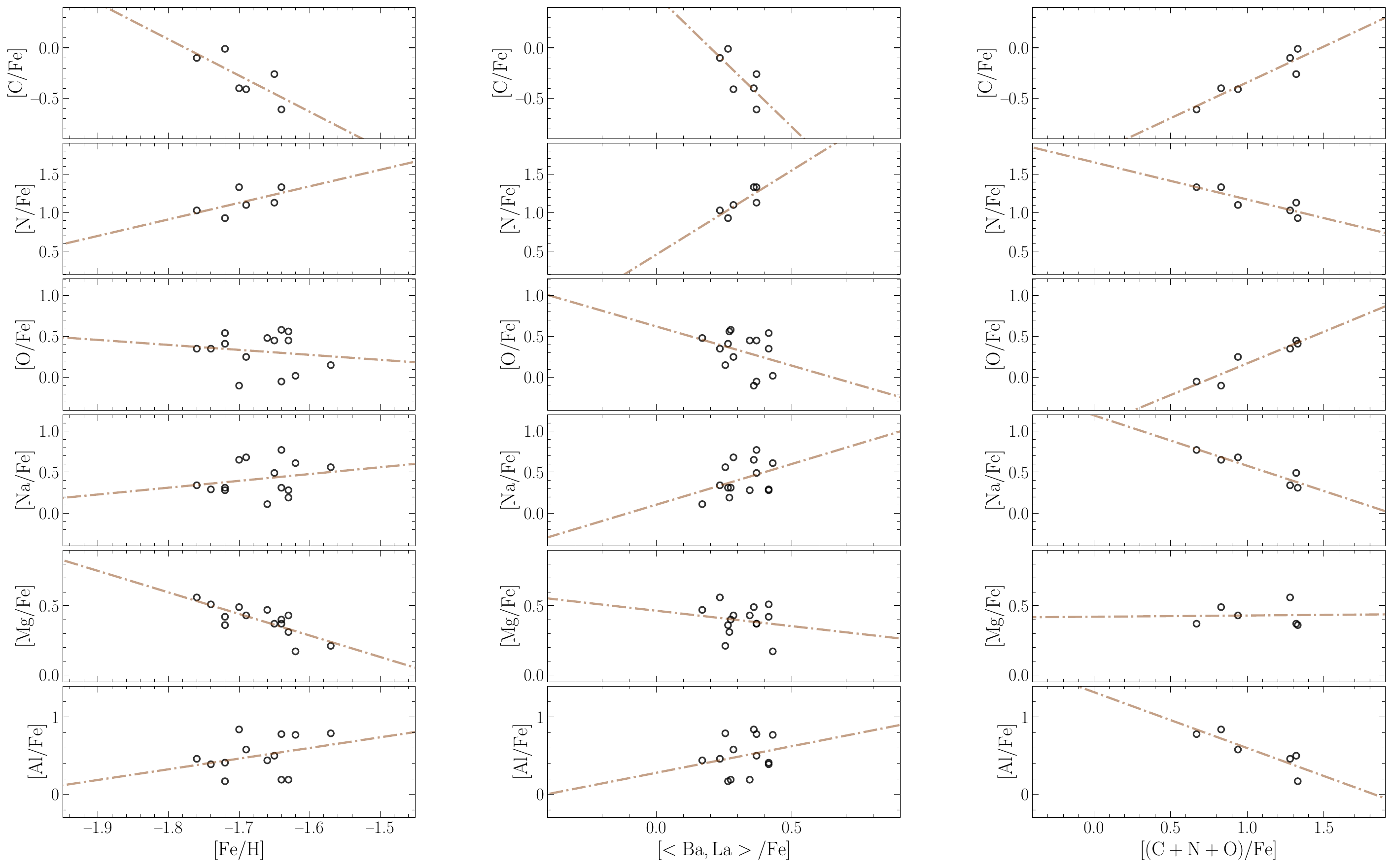}
\caption{
Same as Figure~\ref{fig:trend} but with abundances from \citet{marino2011b}.
}
\label{fig:trend_m2011}
\end{figure*}

Figure~\ref{fig:trend_m2011} presents the abundances of [C/Fe], [N/Fe], [O/Fe], [Na/Fe], [Mg/Fe], and [Al/Fe] as a function of [Fe/H] (left column), the average s-process abundance, $[\langle{\rm Ba,La}\rangle/{\rm Fe}]$, adopted as a proxy for the overall s-process enrichment (middle column), and [(C+N+O)/Fe] (right column). We consider only the anomalous stars classified as s-rich by \citet[][their Tables~5 and~6]{marino2011b}. Brown dot-dashed lines represent the corresponding linear best fits.

Remarkably, the higher-resolution dataset reproduces the same qualitative behavior observed with APOGEE. Carbon, oxygen, and magnesium exhibit decreasing trends with increasing [Fe/H] and $[\langle{\rm Ba,La}\rangle/{\rm Fe}]$, whereas nitrogen and aluminum increase. Conversely, all these trends are reversed when considering the total C+N+O abundance, with the most chemically extreme anomalous stars displaying the lowest [(C+N+O)/Fe] values. Owing to the availability of sodium abundances in the \citet{marino2011b} dataset, we can also extend the comparison to this element. Sodium follows the same behavior as nitrogen and aluminum, increasing toward the most Fe- and s-process-rich anomalous stars, consistently with the interpretation that the chemically most extreme anomalous populations are also the most Na-rich.

Although the trends of individual elements remain relatively modest, their remarkable coherence across all the light-elements, together with the independent confirmation provided by the higher-resolution abundances of \citet{marino2011b}, strongly supports the physical origin of the internal chemical correlations discussed throughout this work.

\bibliography{sample702}{}

@ARTICLE{abdurro2022,
       author = {{Abdurro'uf} and {Accetta}, Katherine and {Aerts}, Conny and {Silva Aguirre}, V{\'\i}ctor and {Ahumada}, Romina and {Ajgaonkar}, Nikhil and {Filiz Ak}, N. and {Alam}, Shadab and {Allende Prieto}, Carlos and {Almeida}, Andr{\'e}s and {Anders}, Friedrich and {Anderson}, Scott F. and {Andrews}, Brett H. and {Anguiano}, Borja and {Aquino-Ort{\'\i}z}, Erik and {Arag{\'o}n-Salamanca}, Alfonso and {Argudo-Fern{\'a}ndez}, Maria and {Ata}, Metin and {Aubert}, Marie and {Avila-Reese}, Vladimir and {Badenes}, Carles and {Barb{\'a}}, Rodolfo H. and {Barger}, Kat and {Barrera-Ballesteros}, Jorge K. and {Beaton}, Rachael L. and {Beers}, Timothy C. and {Belfiore}, Francesco and {Bender}, Chad F. and {Bernardi}, Mariangela and {Bershady}, Matthew A. and {Beutler}, Florian and {Bidin}, Christian Moni and {Bird}, Jonathan C. and {Bizyaev}, Dmitry and {Blanc}, Guillermo A. and {Blanton}, Michael R. and {Boardman}, Nicholas Fraser and {Bolton}, Adam S. and {Boquien}, M{\'e}d{\'e}ric and {Borissova}, Jura and {Bovy}, Jo and {Brandt}, W.~N. and {Brown}, Jordan and {Brownstein}, Joel R. and {Brusa}, Marcella and {Buchner}, Johannes and {Bundy}, Kevin and {Burchett}, Joseph N. and {Bureau}, Martin and {Burgasser}, Adam and {Cabang}, Tuesday K. and {Campbell}, Stephanie and {Cappellari}, Michele and {Carlberg}, Joleen K. and {Wanderley}, F{\'a}bio Carneiro and {Carrera}, Ricardo and {Cash}, Jennifer and {Chen}, Yan-Ping and {Chen}, Wei-Huai and {Cherinka}, Brian and {Chiappini}, Cristina and {Choi}, Peter Doohyun and {Chojnowski}, S. Drew and {Chung}, Haeun and {Clerc}, Nicolas and {Cohen}, Roger E. and {Comerford}, Julia M. and {Comparat}, Johan and {da Costa}, Luiz and {Covey}, Kevin and {Crane}, Jeffrey D. and {Cruz-Gonzalez}, Irene and {Culhane}, Connor and {Cunha}, Katia and {Dai}, Y. Sophia and {Damke}, Guillermo and {Darling}, Jeremy and {Davidson}, James W., Jr. and {Davies}, Roger and {Dawson}, Kyle and {De Lee}, Nathan and {Diamond-Stanic}, Aleksandar M. and {Cano-D{\'\i}az}, Mariana and {S{\'a}nchez}, Helena Dom{\'\i}nguez and {Donor}, John and {Duckworth}, Chris and {Dwelly}, Tom and {Eisenstein}, Daniel J. and {Elsworth}, Yvonne P. and {Emsellem}, Eric and {Eracleous}, Mike and {Escoffier}, Stephanie and {Fan}, Xiaohui and {Farr}, Emily and {Feng}, Shuai and {Fern{\'a}ndez-Trincado}, Jos{\'e} G. and {Feuillet}, Diane and {Filipp}, Andreas and {Fillingham}, Sean P. and {Frinchaboy}, Peter M. and {Fromenteau}, Sebastien and {Galbany}, Llu{\'\i}s and {Garc{\'\i}a}, Rafael A. and {Garc{\'\i}a-Hern{\'a}ndez}, D.~A. and {Ge}, Junqiang and {Geisler}, Doug and {Gelfand}, Joseph and {G{\'e}ron}, Tobias and {Gibson}, Benjamin J. and {Goddy}, Julian and {Godoy-Rivera}, Diego and {Grabowski}, Kathleen and {Green}, Paul J. and {Greener}, Michael and {Grier}, Catherine J. and {Griffith}, Emily and {Guo}, Hong and {Guy}, Julien and {Hadjara}, Massinissa and {Harding}, Paul and {Hasselquist}, Sten and {Hayes}, Christian R. and {Hearty}, Fred and {Hern{\'a}ndez}, Jes{\'u}s and {Hill}, Lewis and {Hogg}, David W. and {Holtzman}, Jon A. and {Horta}, Danny and {Hsieh}, Bau-Ching and {Hsu}, Chin-Hao and {Hsu}, Yun-Hsin and {Huber}, Daniel and {Huertas-Company}, Marc and {Hutchinson}, Brian and {Hwang}, Ho Seong and {Ibarra-Medel}, H{\'e}ctor J. and {Chitham}, Jacob Ider and {Ilha}, Gabriele S. and {Imig}, Julie and {Jaekle}, Will and {Jayasinghe}, Tharindu and {Ji}, Xihan and {Johnson}, Jennifer A. and {Jones}, Amy and {J{\"o}nsson}, Henrik and {Katkov}, Ivan and {Khalatyan}, Arman, Dr. and {Kinemuchi}, Karen and {Kisku}, Shobhit and {Knapen}, Johan H. and {Kneib}, Jean-Paul and {Kollmeier}, Juna A. and {Kong}, Miranda and {Kounkel}, Marina and {Kreckel}, Kathryn and {Krishnarao}, Dhanesh and {Lacerna}, Ivan and {Lane}, Richard R. and {Langgin}, Rachel and {Lavender}, Ramon and {Law}, David R. and {Lazarz}, Daniel and {Leung}, Henry W. and {Leung}, Ho-Hin and {Lewis}, Hannah M. and {Li}, Cheng and {Li}, Ran and {Lian}, Jianhui and {Liang}, Fu-Heng and {Lin}, Lihwai and {Lin}, Yen-Ting and {Lin}, Sicheng and {Lintott}, Chris and {Long}, Dan and {Longa-Pe{\~n}a}, Pen{\'e}lope and {L{\'o}pez-Cob{\'a}}, Carlos and {Lu}, Shengdong and {Lundgren}, Britt F. and {Luo}, Yuanze and {Mackereth}, J. Ted and {de la Macorra}, Axel and {Mahadevan}, Suvrath and {Majewski}, Steven R. and {Manchado}, Arturo and {Mandeville}, Travis and {Maraston}, Claudia and {Margalef-Bentabol}, Berta and {Masseron}, Thomas and {Masters}, Karen L. and {Mathur}, Savita and {McDermid}, Richard M. and {Mckay}, Myles and {Merloni}, Andrea and {Merrifield}, Michael and {Meszaros}, Szabolcs and {Miglio}, Andrea and {Di Mille}, Francesco and {Minniti}, Dante and {Minsley}, Rebecca and {Monachesi}, Antonela and {Moon}, Jeongin and {Mosser}, Benoit and {Mulchaey}, John and {Muna}, Demitri and {Mu{\~n}oz}, Ricardo R. and {Myers}, Adam D. and {Myers}, Natalie and {Nadathur}, Seshadri and {Nair}, Preethi and {Nandra}, Kirpal and {Neumann}, Justus and {Newman}, Jeffrey A. and {Nidever}, David L. and {Nikakhtar}, Farnik and {Nitschelm}, Christian and {O'Connell}, Julia E. and {Garma-Oehmichen}, Luis and {Luan Souza de Oliveira}, Gabriel and {Olney}, Richard and {Oravetz}, Daniel and {Ortigoza-Urdaneta}, Mario and {Osorio}, Yeisson and {Otter}, Justin and {Pace}, Zachary J. and {Padilla}, Nelson and {Pan}, Kaike and {Pan}, Hsi-An and {Parikh}, Taniya and {Parker}, James and {Peirani}, Sebastien and {Pe{\~n}a Ram{\'\i}rez}, Karla and {Penny}, Samantha and {Percival}, Will J. and {Perez-Fournon}, Ismael and {Pinsonneault}, Marc and {Poidevin}, Fr{\'e}d{\'e}rick and {Poovelil}, Vijith Jacob and {Price-Whelan}, Adrian M. and {B{\'a}rbara de Andrade Queiroz}, Anna and {Raddick}, M. Jordan and {Ray}, Amy and {Rembold}, Sandro Barboza and {Riddle}, Nicole and {Riffel}, Rogemar A. and {Riffel}, Rog{\'e}rio and {Rix}, Hans-Walter and {Robin}, Annie C. and {Rodr{\'\i}guez-Puebla}, Aldo and {Roman-Lopes}, Alexandre and {Rom{\'a}n-Z{\'u}{\~n}iga}, Carlos and {Rose}, Benjamin and {Ross}, Ashley J. and {Rossi}, Graziano and {Rubin}, Kate H.~R. and {Salvato}, Mara and {S{\'a}nchez}, Seb{\'a}stian F. and {S{\'a}nchez-Gallego}, Jos{\'e} R. and {Sanderson}, Robyn and {Santana Rojas}, Felipe Antonio and {Sarceno}, Edgar and {Sarmiento}, Regina and {Sayres}, Conor and {Sazonova}, Elizaveta and {Schaefer}, Adam L. and {Schiavon}, Ricardo and {Schlegel}, David J. and {Schneider}, Donald P. and {Schultheis}, Mathias and {Schwope}, Axel and {Serenelli}, Aldo and {Serna}, Javier and {Shao}, Zhengyi and {Shapiro}, Griffin and {Sharma}, Anubhav and {Shen}, Yue and {Shetrone}, Matthew and {Shu}, Yiping and {Simon}, Joshua D. and {Skrutskie}, M.~F. and {Smethurst}, Rebecca and {Smith}, Verne and {Sobeck}, Jennifer and {Spoo}, Taylor and {Sprague}, Dani and {Stark}, David V. and {Stassun}, Keivan G. and {Steinmetz}, Matthias and {Stello}, Dennis and {Stone-Martinez}, Alexander and {Storchi-Bergmann}, Thaisa and {Stringfellow}, Guy S. and {Stutz}, Amelia and {Su}, Yung-Chau and {Taghizadeh-Popp}, Manuchehr and {Talbot}, Michael S. and {Tayar}, Jamie and {Telles}, Eduardo and {Teske}, Johanna and {Thakar}, Ani and {Theissen}, Christopher and {Tkachenko}, Andrew and {Thomas}, Daniel and {Tojeiro}, Rita and {Hernandez Toledo}, Hector and {Troup}, Nicholas W. and {Trump}, Jonathan R. and {Trussler}, James and {Turner}, Jacqueline and {Tuttle}, Sarah and {Unda-Sanzana}, Eduardo and {V{\'a}zquez-Mata}, Jos{\'e} Antonio and {Valentini}, Marica and {Valenzuela}, Octavio and {Vargas-Gonz{\'a}lez}, Jaime and {Vargas-Maga{\~n}a}, Mariana and {Alfaro}, Pablo Vera and {Villanova}, Sandro and {Vincenzo}, Fiorenzo and {Wake}, David and {Warfield}, Jack T. and {Washington}, Jessica Diane and {Weaver}, Benjamin Alan and {Weijmans}, Anne-Marie and {Weinberg}, David H. and {Weiss}, Achim and {Westfall}, Kyle B. and {Wild}, Vivienne and {Wilde}, Matthew C. and {Wilson}, John C. and {Wilson}, Robert F. and {Wilson}, Mikayla and {Wolf}, Julien and {Wood-Vasey}, W.~M. and {Yan}, Renbin and {Zamora}, Olga and {Zasowski}, Gail and {Zhang}, Kai and {Zhao}, Cheng and {Zheng}, Zheng and {Zheng}, Zheng and {Zhu}, Kai},
       title = "{The Seventeenth Data Release of the Sloan Digital Sky Surveys: Complete Release of MaNGA, MaStar, and APOGEE-2 Data}",
      journal = {\apjs},
         year = 2022,
        month = apr,
       volume = {259},
       number = {2},
          eid = {35},
        pages = {35},
          doi = {10.3847/1538-4365/ac4414},
archivePrefix = {arXiv},
       eprint = {2112.02026},
 primaryClass = {astro-ph.GA},
       adsurl = {https://ui.adsabs.harvard.edu/abs/2022ApJS..259...35A}
}

@ARTICLE{alvesbrito2012,
       author = {{Alves-Brito}, A. and {Yong}, D. and {Mel{\'e}ndez}, J. and {V{\'a}squez}, S. and {Karakas}, A.~I.},
        title = "{CNO and F abundances in the globular cluster M 22 (NGC 6656)}",
      journal = {\aap},
         year = 2012,
        month = apr,
       volume = {540},
          eid = {A3},
        pages = {A3},
          doi = {10.1051/0004-6361/201118623},
archivePrefix = {arXiv},
       eprint = {1202.0797},
 primaryClass = {astro-ph.SR},
       adsurl = {https://ui.adsabs.harvard.edu/abs/2012A&A...540A...3A}
}

@ARTICLE{anderson2000,
       author = {{Anderson}, Jay and {King}, Ivan R.},
        title = "{Toward High-Precision Astrometry with WFPC2. I. Deriving an Accurate Point-Spread Function}",
      journal = {\pasp},
         year = 2000,
        month = oct,
       volume = {112},
       number = {776},
        pages = {1360-1382},
          doi = {10.1086/316632},
archivePrefix = {arXiv},
       eprint = {astro-ph/0006325},
 primaryClass = {astro-ph},
       adsurl = {https://ui.adsabs.harvard.edu/abs/2000PASP..112.1360A}
}

@ARTICLE{anderson2006,
       author = {{Anderson}, J. and {Bedin}, L.~R. and {Piotto}, G. and {Yadav}, R.~S. and {Bellini}, A.},
        title = "{Ground-based CCD astrometry with wide field imagers. I. Observations just a few years apart allow decontamination of field objects from members in two globular clusters}",
      journal = {\aap},
         year = 2006,
        month = aug,
       volume = {454},
       number = {3},
        pages = {1029-1045},
          doi = {10.1051/0004-6361:20065004},
archivePrefix = {arXiv},
       eprint = {astro-ph/0604541},
 primaryClass = {astro-ph},
       adsurl = {https://ui.adsabs.harvard.edu/abs/2006A&A...454.1029A}
}

@ARTICLE{anderson2008,
       author = {{Anderson}, Jay and {Sarajedini}, Ata and {Bedin}, Luigi R. and {King}, Ivan R. and {Piotto}, Giampaolo and {Reid}, I. Neill and {Siegel}, Michael and {Majewski}, Steven R. and {Paust}, Nathaniel E.~Q. and {Aparicio}, Antonio and {Milone}, Antonino P. and {Chaboyer}, Brian and {Rosenberg}, Alfred},
        title = "{The Acs Survey of Globular Clusters. V. Generating a Comprehensive Star Catalog for each Cluster}",
      journal = {\aj},
         year = 2008,
        month = jun,
       volume = {135},
       number = {6},
        pages = {2055-2073},
          doi = {10.1088/0004-6256/135/6/2055},
archivePrefix = {arXiv},
       eprint = {0804.2025},
 primaryClass = {astro-ph},
       adsurl = {https://ui.adsabs.harvard.edu/abs/2008AJ....135.2055A}
}

@ARTICLE{bastian2018,
       author = {{Bastian}, Nate and {Lardo}, Carmela},
        title = "{Multiple Stellar Populations in Globular Clusters}",
      journal = {\araa},
         year = 2018,
        month = sep,
       volume = {56},
        pages = {83-136},
          doi = {10.1146/annurev-astro-081817-051839},
archivePrefix = {arXiv},
       eprint = {1712.01286},
 primaryClass = {astro-ph.SR},
       adsurl = {https://ui.adsabs.harvard.edu/abs/2018ARA&A..56...83B}
}

@ARTICLE{bekki2016,
       author = {{Bekki}, Kenji and {Tsujimoto}, Takuji},
        title = "{Formation of Anomalous Globular Clusters with Metallicity Spreads: A Unified Picture}",
      journal = {\apj},
         year = 2016,
        month = nov,
       volume = {831},
       number = {1},
          eid = {70},
        pages = {70},
          doi = {10.3847/0004-637X/831/1/70},
       adsurl = {https://ui.adsabs.harvard.edu/abs/2016ApJ...831...70B}
}

@ARTICLE{bekki2026,
       author = {{Bekki}, Kenji and {McKenzie}, Madeleine},
        title = "{Globular cluster formation with multiple stellar populations: A comprehensive overview of a star-cloud interaction scenario}",
      journal = {arXiv e-prints},
         year = 2026,
        month = jun,
          eid = {arXiv:2606.29707},
        pages = {arXiv:2606.29707},
          doi = {10.48550/arXiv.2606.29707},
archivePrefix = {arXiv},
       eprint = {2606.29707},
 primaryClass = {astro-ph.GA},
       adsurl = {https://ui.adsabs.harvard.edu/abs/2026arXiv260629707B}
}

@ARTICLE{bellazzini2008,
       author = {{Bellazzini}, M. and {Ibata}, R.~A. and {Chapman}, S.~C. and {Mackey}, A.~D. and {Monaco}, L. and {Irwin}, M.~J. and {Martin}, N.~F. and {Lewis}, G.~F. and {Dalessandro}, E.},
        title = "{The Nucleus of the Sagittarius Dsph Galaxy and M54: a Window on the Process of Galaxy Nucleation}",
      journal = {\aj},
         year = 2008,
        month = sep,
       volume = {136},
       number = {3},
        pages = {1147-1170},
          doi = {10.1088/0004-6256/136/3/1147},
archivePrefix = {arXiv},
       eprint = {0807.0105},
 primaryClass = {astro-ph},
       adsurl = {https://ui.adsabs.harvard.edu/abs/2008AJ....136.1147B}
}

@ARTICLE{carretta2010,
       author = {{Carretta}, E. and {Bragaglia}, A. and {Gratton}, R.~G. and {Recio-Blanco}, A. and {Lucatello}, S. and {D'Orazi}, V. and {Cassisi}, S.},
        title = "{Properties of stellar generations in globular clusters and relations with global parameters}",
      journal = {\aap},
         year = 2010,
        month = jun,
       volume = {516},
          eid = {A55},
        pages = {A55},
          doi = {10.1051/0004-6361/200913451},
archivePrefix = {arXiv},
       eprint = {1003.1723},
 primaryClass = {astro-ph.GA},
       adsurl = {https://ui.adsabs.harvard.edu/abs/2010A&A...516A..55C}
}

@ARTICLE{carretta2011,
       author = {{Carretta}, E. and {Lucatello}, S. and {Gratton}, R.~G. and {Bragaglia}, A. and {D'Orazi}, V.},
        title = "{Multiple stellar populations in the globular cluster NGC 1851}",
      journal = {\aap},
         year = 2011,
        month = sep,
       volume = {533},
          eid = {A69},
        pages = {A69},
          doi = {10.1051/0004-6361/201117269},
archivePrefix = {arXiv},
       eprint = {1106.3174},
 primaryClass = {astro-ph.SR},
       adsurl = {https://ui.adsabs.harvard.edu/abs/2011A&A...533A..69C}
}

@ARTICLE{catelan2009,
       author = {{Catelan}, M.},
        title = "{Horizontal branch stars: the interplay between observations and theory, and insights into the formation of the Galaxy}",
      journal = {\apss},
         year = 2009,
        month = apr,
       volume = {320},
       number = {4},
        pages = {261-309},
          doi = {10.1007/s10509-009-9987-8},
archivePrefix = {arXiv},
       eprint = {astro-ph/0507464},
 primaryClass = {astro-ph},
       adsurl = {https://ui.adsabs.harvard.edu/abs/2009Ap&SS.320..261C}
}

@ARTICLE{cordoni2018,
       author = {{Cordoni}, G. and {Milone}, A.~P. and {Marino}, A.~F. and {Di Criscienzo}, M. and {D'Antona}, F. and {Dotter}, A. and {Lagioia}, E.~P. and {Tailo}, M.},
        title = "{Extended Main-sequence Turnoff as a Common Feature of Milky Way Open Clusters}",
      journal = {\apj},
         year = 2018,
        month = dec,
       volume = {869},
       number = {2},
          eid = {139},
        pages = {139},
          doi = {10.3847/1538-4357/aaedc1},
archivePrefix = {arXiv},
       eprint = {1811.01192},
 primaryClass = {astro-ph.SR},
       adsurl = {https://ui.adsabs.harvard.edu/abs/2018ApJ...869..139C}
}

@ARTICLE{cristallo2009,
       author = {{Cristallo}, S. and {Straniero}, O. and {Gallino}, R. and {Piersanti}, L. and {Dom{\'\i}nguez}, I. and {Lederer}, M.~T.},
        title = "{Evolution, Nucleosynthesis, and Yields of Low-Mass Asymptotic Giant Branch Stars at Different Metallicities}",
      journal = {\apj},
         year = 2009,
        month = may,
       volume = {696},
       number = {1},
        pages = {797-820},
          doi = {10.1088/0004-637X/696/1/797},
archivePrefix = {arXiv},
       eprint = {0902.0243},
 primaryClass = {astro-ph.SR},
       adsurl = {https://ui.adsabs.harvard.edu/abs/2009ApJ...696..797C}
}

@ARTICLE{dacosta2009,
       author = {{Da Costa}, G.~S. and {Held}, E.~V. and {Saviane}, I. and {Gullieuszik}, M.},
        title = "{M22: An [Fe/H] Abundance Range Revealed}",
      journal = {\apj},
         year = 2009,
        month = nov,
       volume = {705},
       number = {2},
        pages = {1481-1491},
          doi = {10.1088/0004-637X/705/2/1481},
archivePrefix = {arXiv},
       eprint = {0909.5265},
 primaryClass = {astro-ph.GA},
       adsurl = {https://ui.adsabs.harvard.edu/abs/2009ApJ...705.1481D}
}

@ARTICLE{dacosta2011,
       author = {{Da Costa}, G.~S. and {Marino}, A.~F.},
        title = "{Nucleosynthesis in the Stellar Systems {\ensuremath{\omega}} Centauri and M22}",
      journal = {\pasa},
         year = 2011,
        month = jan,
       volume = {28},
       number = {1},
        pages = {28-37},
          doi = {10.1071/AS10027},
archivePrefix = {arXiv},
       eprint = {1009.1955},
 primaryClass = {astro-ph.SR},
       adsurl = {https://ui.adsabs.harvard.edu/abs/2011PASA...28...28D}
}

@ARTICLE{dantona2016,
       author = {{D'Antona}, F. and {Vesperini}, E. and {D'Ercole}, A. and {Ventura}, P. and {Milone}, A.~P. and {Marino}, A.~F. and {Tailo}, M.},
        title = "{A single model for the variety of multiple-population formation(s) in globular clusters: a temporal sequence}",
      journal = {\mnras},
         year = 2016,
        month = may,
       volume = {458},
       number = {2},
        pages = {2122-2139},
          doi = {10.1093/mnras/stw387},
archivePrefix = {arXiv},
       eprint = {1602.05412},
 primaryClass = {astro-ph.GA},
       adsurl = {https://ui.adsabs.harvard.edu/abs/2016MNRAS.458.2122D}
}

@ARTICLE{dondoglio2023,
       author = {{Dondoglio}, E. and {Milone}, A.~P. and {Marino}, A.~F. and {D'Antona}, F. and {Cordoni}, G. and {Legnardi}, M.~V. and {Lagioia}, E.~P. and {Jang}, S. and {Ziliotto}, T. and {Carlos}, M. and {Dell'Agli}, F. and {Karakas}, A. and {Mohandasan}, A. and {Osborn}, Z. and {Tailo}, M. and {Ventura}, P.},
        title = "{A deep dive into the Type II globular cluster NGC 1851}",
      journal = {\mnras},
         year = 2023,
        month = dec,
       volume = {526},
       number = {2},
        pages = {2960-2976},
          doi = {10.1093/mnras/stad2950},
archivePrefix = {arXiv},
       eprint = {2309.16423},
 primaryClass = {astro-ph.GA},
       adsurl = {https://ui.adsabs.harvard.edu/abs/2023MNRAS.526.2960D}
}

@ARTICLE{dondoglio2025,
       author = {{Dondoglio}, E. and {Marino}, A.~F. and {Milone}, A.~P. and {Jang}, S. and {Cordoni}, G. and {D'Antona}, F. and {Renzini}, A. and {Tailo}, M. and {Sanchez}, A. Bouras Moreno and {Muratore}, F. and {Ziliotto}, T. and {Barbieri}, M. and {Bortolan}, E. and {Lagioia}, E.~P. and {Legnardi}, M.~V. and {Lionetto}, S. and {Mohandasan}, A.},
        title = "{Linking photometry and spectroscopy: profiling multiple populations in globular clusters}",
      journal = {\aap},
         year = 2025,
        month = may,
       volume = {697},
          eid = {A135},
        pages = {A135},
          doi = {10.1051/0004-6361/202453024},
archivePrefix = {arXiv},
       eprint = {2503.15976},
 primaryClass = {astro-ph.GA},
       adsurl = {https://ui.adsabs.harvard.edu/abs/2025A&A...697A.135D}
}

@ARTICLE{dondoglio2026,
       author = {{Dondoglio}, E. and {Milone}, A.~P. and {Marino}, A.~F. and {Mastrobuono-Battisti}, A. and {Bortolan}, E. and {Legnardi}, M.~V. and {Ziliotto}, T. and {Muratore}, F. and {Cordoni}, G. and {Lagioia}, E.~P. and {Tailo}, M.},
        title = "{Tracing {\ensuremath{\omega}}Centauri's origins: Spatial and chemical signatures of its formation history}",
      journal = {\aap},
         year = 2026,
        month = jan,
       volume = {705},
          eid = {A2},
        pages = {A2},
          doi = {10.1051/0004-6361/202556545},
archivePrefix = {arXiv},
       eprint = {2509.16719},
 primaryClass = {astro-ph.GA},
       adsurl = {https://ui.adsabs.harvard.edu/abs/2026A&A...705A...2D}
}

@ARTICLE{dondoglio2026b,
       author = {{Dondoglio}, Emanuele and {Milone}, A.~P. and {Marino}, A.~F. and {Cordoni}, G. and {Legnardi}, M.~V. and {Ziliotto}, T. and {Asa'd}, R. and {Mastrobuono-Battisti}, A. and {Muratore}, F. and {Bortolan}, E. and {Lagioia}, E.~P. and {Tailo}, M.},
        title = "{Multiple populations along the asymptotic giant branch: a Gaia+APOGEE study of 22 Galactic globular clusters}",
      journal = {arXiv e-prints},
         year = 2026,
        month = jul,
          eid = {arXiv:2607.08376},
        pages = {arXiv:2607.08376},
          doi = {10.48550/arXiv.2607.08376},
archivePrefix = {arXiv},
       eprint = {2607.08376},
 primaryClass = {astro-ph.GA},
       adsurl = {https://ui.adsabs.harvard.edu/abs/2026arXiv260708376D}
}

@ARTICLE{fernandextrincado2021,
       author = {{Fern{\'a}ndez-Trincado}, Jos{\'e} G. and {Beers}, Timothy C. and {Barbuy}, Beatriz and {M{\'e}sz{\'a}ros}, Szabolcs and {Minniti}, Dante and {Smith}, Verne V. and {Cunha}, Katia and {Villanova}, Sandro and {Geisler}, Doug and {Majewski}, Steven R. and {Carigi}, Leticia and {Tang}, Baitian and {Moni Bidin}, Christian and {Vieira}, Katherine},
        title = "{APOGEE-2S Discovery of Light- and Heavy-element Abundance Correlations in the Bulge Globular Cluster NGC 6380}",
      journal = {\apjl},
         year = 2021,
        month = sep,
       volume = {918},
       number = {1},
          eid = {L9},
        pages = {L9},
          doi = {10.3847/2041-8213/ac1c7e},
archivePrefix = {arXiv},
       eprint = {2109.02661},
 primaryClass = {astro-ph.GA},
       adsurl = {https://ui.adsabs.harvard.edu/abs/2021ApJ...918L...9F}
}

@ARTICLE{fernandeztrincado2022,
       author = {{Fern{\'a}ndez-Trincado}, Jos{\'e} G. and {Villanova}, Sandro and {Geisler}, Doug and {Barbuy}, Beatriz and {Minniti}, Dante and {Beers}, Timothy C. and {M{\'e}sz{\'a}ros}, Szabolcs and {Tang}, Baitian and {Cohen}, Roger E. and {Moni Bidin}, Cristian and {Garro}, Elisa R. and {Baeza}, Ian and {Mu{\~n}oz}, Cesar},
        title = "{CAPOS: The bulge Cluster APOgee Survey. III. Spectroscopic tomography of Tonantzintla 2}",
      journal = {\aap},
         year = 2022,
        month = feb,
       volume = {658},
          eid = {A116},
        pages = {A116},
          doi = {10.1051/0004-6361/202141742},
archivePrefix = {arXiv},
       eprint = {2110.10700},
 primaryClass = {astro-ph.GA},
       adsurl = {https://ui.adsabs.harvard.edu/abs/2022A&A...658A.116F}
}

@ARTICLE{gaia2023,
       author = {{Gaia Collaboration} and {Vallenari}, A. and {Brown}, A.~G.~A. and {Prusti}, T. and {de Bruijne}, J.~H.~J. and {Arenou}, F. and {Babusiaux}, C. and {Biermann}, M. and {Creevey}, O.~L. and {Ducourant}, C. and {Evans}, D.~W. and {Eyer}, L. and {Guerra}, R. and {Hutton}, A. and {Jordi}, C. and {Klioner}, S.~A. and {Lammers}, U.~L. and {Lindegren}, L. and {Luri}, X. and {Mignard}, F. and {Panem}, C. and {Pourbaix}, D. and {Randich}, S. and {Sartoretti}, P. and {Soubiran}, C. and {Tanga}, P. and {Walton}, N.~A. and {Bailer-Jones}, C.~A.~L. and {Bastian}, U. and {Drimmel}, R. and {Jansen}, F. and {Katz}, D. and {Lattanzi}, M.~G. and {van Leeuwen}, F. and {Bakker}, J. and {Cacciari}, C. and {Casta{\~n}eda}, J. and {De Angeli}, F. and {Fabricius}, C. and {Fouesneau}, M. and {Fr{\'e}mat}, Y. and {Galluccio}, L. and {Guerrier}, A. and {Heiter}, U. and {Masana}, E. and {Messineo}, R. and {Mowlavi}, N. and {Nicolas}, C. and {Nienartowicz}, K. and {Pailler}, F. and {Panuzzo}, P. and {Riclet}, F. and {Roux}, W. and {Seabroke}, G.~M. and {Sordo}, R. and {Th{\'e}venin}, F. and {Gracia-Abril}, G. and {Portell}, J. and {Teyssier}, D. and {Altmann}, M. and {Andrae}, R. and {Audard}, M. and {Bellas-Velidis}, I. and {Benson}, K. and {Berthier}, J. and {Blomme}, R. and {Burgess}, P.~W. and {Busonero}, D. and {Busso}, G. and {C{\'a}novas}, H. and {Carry}, B. and {Cellino}, A. and {Cheek}, N. and {Clementini}, G. and {Damerdji}, Y. and {Davidson}, M. and {de Teodoro}, P. and {Nu{\~n}ez Campos}, M. and {Delchambre}, L. and {Dell'Oro}, A. and {Esquej}, P. and {Fern{\'a}ndez-Hern{\'a}ndez}, J. and {Fraile}, E. and {Garabato}, D. and {Garc{\'\i}a-Lario}, P. and {Gosset}, E. and {Haigron}, R. and {Halbwachs}, J. -L. and {Hambly}, N.~C. and {Harrison}, D.~L. and {Hern{\'a}ndez}, J. and {Hestroffer}, D. and {Hodgkin}, S.~T. and {Holl}, B. and {Jan{\ss}en}, K. and {Jevardat de Fombelle}, G. and {Jordan}, S. and {Krone-Martins}, A. and {Lanzafame}, A.~C. and {L{\"o}ffler}, W. and {Marchal}, O. and {Marrese}, P.~M. and {Moitinho}, A. and {Muinonen}, K. and {Osborne}, P. and {Pancino}, E. and {Pauwels}, T. and {Recio-Blanco}, A. and {Reyl{\'e}}, C. and {Riello}, M. and {Rimoldini}, L. and {Roegiers}, T. and {Rybizki}, J. and {Sarro}, L.~M. and {Siopis}, C. and {Smith}, M. and {Sozzetti}, A. and {Utrilla}, E. and {van Leeuwen}, M. and {Abbas}, U. and {{\'A}brah{\'a}m}, P. and {Abreu Aramburu}, A. and {Aerts}, C. and {Aguado}, J.~J. and {Ajaj}, M. and {Aldea-Montero}, F. and {Altavilla}, G. and {{\'A}lvarez}, M.~A. and {Alves}, J. and {Anders}, F. and {Anderson}, R.~I. and {Anglada Varela}, E. and {Antoja}, T. and {Baines}, D. and {Baker}, S.~G. and {Balaguer-N{\'u}{\~n}ez}, L. and {Balbinot}, E. and {Balog}, Z. and {Barache}, C. and {Barbato}, D. and {Barros}, M. and {Barstow}, M.~A. and {Bartolom{\'e}}, S. and {Bassilana}, J. -L. and {Bauchet}, N. and {Becciani}, U. and {Bellazzini}, M. and {Berihuete}, A. and {Bernet}, M. and {Bertone}, S. and {Bianchi}, L. and {Binnenfeld}, A. and {Blanco-Cuaresma}, S. and {Blazere}, A. and {Boch}, T. and {Bombrun}, A. and {Bossini}, D. and {Bouquillon}, S. and {Bragaglia}, A. and {Bramante}, L. and {Breedt}, E. and {Bressan}, A. and {Brouillet}, N. and {Brugaletta}, E. and {Bucciarelli}, B. and {Burlacu}, A. and {Butkevich}, A.~G. and {Buzzi}, R. and {Caffau}, E. and {Cancelliere}, R. and {Cantat-Gaudin}, T. and {Carballo}, R. and {Carlucci}, T. and {Carnerero}, M.~I. and {Carrasco}, J.~M. and {Casamiquela}, L. and {Castellani}, M. and {Castro-Ginard}, A. and {Chaoul}, L. and {Charlot}, P. and {Chemin}, L. and {Chiaramida}, V. and {Chiavassa}, A. and {Chornay}, N. and {Comoretto}, G. and {Contursi}, G. and {Cooper}, W.~J. and {Cornez}, T. and {Cowell}, S. and {Crifo}, F. and {Cropper}, M. and {Crosta}, M. and {Crowley}, C. and {Dafonte}, C. and {Dapergolas}, A. and {David}, M. and {David}, P. and {de Laverny}, P. and {De Luise}, F. and {De March}, R.},
        title = "{Gaia Data Release 3. Summary of the content and survey properties}",
      journal = {\aap},
         year = 2023,
        month = jun,
       volume = {674},
          eid = {A1},
        pages = {A1},
          doi = {10.1051/0004-6361/202243940},
archivePrefix = {arXiv},
       eprint = {2208.00211},
 primaryClass = {astro-ph.GA},
       adsurl = {https://ui.adsabs.harvard.edu/abs/2023A&A...674A...1G}
}

@ARTICLE{gieles2025,
       author = {{Gieles}, Mark and {Padoan}, Paolo and {Charbonnel}, Corinne and {Vink}, Jorick S. and {Ram{\'\i}rez-Galeano}, Laura},
        title = "{Globular cluster formation from inertial inflows: accreting extremely massive stars as the origin of abundance anomalies}",
      journal = {\mnras},
         year = 2025,
        month = nov,
       volume = {544},
       number = {1},
        pages = {483-512},
          doi = {10.1093/mnras/staf1314},
archivePrefix = {arXiv},
       eprint = {2501.12138},
 primaryClass = {astro-ph.GA},
       adsurl = {https://ui.adsabs.harvard.edu/abs/2025MNRAS.544..483G}
}

@ARTICLE{gratton2011,
       author = {{Gratton}, R.~G. and {Lucatello}, S. and {Carretta}, E. and {Bragaglia}, A. and {D'Orazi}, V. and {Momany}, Y. Al},
        title = "{The Na-O anticorrelation in horizontal branch stars. I. NGC 2808}",
      journal = {\aap},
         year = 2011,
        month = oct,
       volume = {534},
          eid = {A123},
        pages = {A123},
          doi = {10.1051/0004-6361/201117690},
archivePrefix = {arXiv},
       eprint = {1109.4013},
 primaryClass = {astro-ph.GA},
       adsurl = {https://ui.adsabs.harvard.edu/abs/2011A&A...534A.123G}
}

@ARTICLE{gratton2019,
       author = {{Gratton}, Raffaele and {Bragaglia}, Angela and {Carretta}, Eugenio and {D'Orazi}, Valentina and {Lucatello}, Sara and {Sollima}, Antonio},
        title = "{What is a globular cluster? An observational perspective}",
      journal = {\aapr},
         year = 2019,
        month = nov,
       volume = {27},
       number = {1},
          eid = {8},
        pages = {8},
          doi = {10.1007/s00159-019-0119-3},
archivePrefix = {arXiv},
       eprint = {1911.02835},
 primaryClass = {astro-ph.SR},
       adsurl = {https://ui.adsabs.harvard.edu/abs/2019A&ARv..27....8G}
}

@ARTICLE{greggio1990,
       author = {{Greggio}, Laura and {Renzini}, Alvio},
        title = "{Clues on the Hot Star Content and the Ultraviolet Output of Elliptical Galaxies}",
      journal = {\apj},
         year = 1990,
        month = nov,
       volume = {364},
        pages = {35},
          doi = {10.1086/169384},
       adsurl = {https://ui.adsabs.harvard.edu/abs/1990ApJ...364...35G}
}

@ARTICLE{grundahl1999,
       author = {{Grundahl}, F. and {Catelan}, M. and {Landsman}, W.~B. and {Stetson}, P.~B. and {Andersen}, M.~I.},
        title = "{Hot Horizontal-Branch Stars: The Ubiquitous Nature of the ``Jump'' in Str{\"o}mgren u, Low Gravities, and the Role of Radiative Levitation of Metals}",
      journal = {\apj},
         year = 1999,
        month = oct,
       volume = {524},
       number = {1},
        pages = {242-261},
          doi = {10.1086/307807},
archivePrefix = {arXiv},
       eprint = {astro-ph/9903120},
 primaryClass = {astro-ph},
       adsurl = {https://ui.adsabs.harvard.edu/abs/1999ApJ...524..242G}
}

@ARTICLE{han2009,
       author = {{Han}, Sang-Il and {Lee}, Young-Wook and {Joo}, Seok-Joo and {Sohn}, Sangmo Tony and {Yoon}, Suk-Jin and {Kim}, Hak-Sub and {Lee}, Jae-Woo},
        title = "{The Presence of Two Distinct Red Giant Branches in the Globular Cluster NGC 1851}",
      journal = {\apjl},
         year = 2009,
        month = dec,
       volume = {707},
       number = {2},
        pages = {L190-L194},
          doi = {10.1088/0004-637X/707/2/L190},
archivePrefix = {arXiv},
       eprint = {0911.5356},
 primaryClass = {astro-ph.GA},
       adsurl = {https://ui.adsabs.harvard.edu/abs/2009ApJ...707L.190H}
}

@ARTICLE{hasselquist2017,
       author = {{Hasselquist}, Sten and {Shetrone}, Matthew and {Smith}, Verne and {Holtzman}, Jon and {McWilliam}, Andrew and {Fern{\'a}ndez-Trincado}, J.~G. and {Beers}, Timothy C. and {Majewski}, Steven R. and {Nidever}, David L. and {Tang}, Baitian and {Tissera}, Patricia B. and {Fern{\'a}ndez Alvar}, Emma and {Allende Prieto}, Carlos and {Almeida}, Andres and {Anguiano}, Borja and {Battaglia}, Giuseppina and {Carigi}, Leticia and {Delgado Inglada}, Gloria and {Frinchaboy}, Peter and {Garc{\'\i}a-Hern{\'a}ndez}, D.~A. and {Geisler}, Doug and {Minniti}, Dante and {Placco}, Vinicius M. and {Schultheis}, Mathias and {Sobeck}, Jennifer and {Villanova}, Sandro},
        title = "{APOGEE Chemical Abundances of the Sagittarius Dwarf Galaxy}",
      journal = {\apj},
         year = 2017,
        month = aug,
       volume = {845},
       number = {2},
          eid = {162},
        pages = {162},
          doi = {10.3847/1538-4357/aa7ddc},
archivePrefix = {arXiv},
       eprint = {1707.03456},
 primaryClass = {astro-ph.GA},
       adsurl = {https://ui.adsabs.harvard.edu/abs/2017ApJ...845..162H}
}

@ARTICLE{harris1996,
       author = {{Harris}, William E.},
        title = "{A Catalog of Parameters for Globular Clusters in the Milky Way}",
      journal = {\aj},
         year = 1996,
        month = oct,
       volume = {112},
        pages = {1487},
          doi = {10.1086/118116},
       adsurl = {https://ui.adsabs.harvard.edu/abs/1996AJ....112.1487H}
}

@ARTICLE{jang2022,
       author = {{Jang}, S. and {Milone}, A.~P. and {Legnardi}, M.~V. and {Marino}, A.~F. and {Mastrobuono-Battisti}, A. and {Dondoglio}, E. and {Lagioia}, E.~P. and {Casagrande}, L. and {Carlos}, M. and {Mohandasan}, A. and {Cordoni}, G. and {Bortolan}, E. and {Lee}, Y. -W.},
        title = "{Chromosome maps of globular clusters from wide-field ground-based photometry}",
      journal = {\mnras},
         year = 2022,
        month = dec,
       volume = {517},
       number = {4},
        pages = {5687-5703},
          doi = {10.1093/mnras/stac3086},
archivePrefix = {arXiv},
       eprint = {2211.00650},
 primaryClass = {astro-ph.GA},
       adsurl = {https://ui.adsabs.harvard.edu/abs/2022MNRAS.517.5687J}
}

@ARTICLE{kobayashi2025,
       author = {{Kobayashi}, Chiaki},
        title = "{Nucleosynthesis and the chemical enrichment of galaxies}",
      journal = {arXiv e-prints},
         year = 2025,
        month = jun,
          eid = {arXiv:2506.20436},
        pages = {arXiv:2506.20436},
          doi = {10.48550/arXiv.2506.20436},
archivePrefix = {arXiv},
       eprint = {2506.20436},
 primaryClass = {astro-ph.GA},
       adsurl = {https://ui.adsabs.harvard.edu/abs/2025arXiv250620436K}
}

@ARTICLE{landolt1992,
       author = {{Landolt}, Arlo U.},
        title = "{UBVRI Photometric Standard Stars in the Magnitude Range 11.5 < V < 16.0 Around the Celestial Equator}",
      journal = {\aj},
         year = 1992,
        month = jul,
       volume = {104},
        pages = {340},
          doi = {10.1086/116242},
       adsurl = {https://ui.adsabs.harvard.edu/abs/1992AJ....104..340L}
}

@ARTICLE{lee2009,
       author = {{Lee}, Jae-Woo and {Kang}, Young-Woon and {Lee}, Jina and {Lee}, Young-Wook},
        title = "{Enrichment by supernovae in globular clusters with multiple populations}",
      journal = {\nat},
         year = 2009,
        month = nov,
       volume = {462},
       number = {7272},
        pages = {480-482},
          doi = {10.1038/nature08565},
archivePrefix = {arXiv},
       eprint = {0911.4798},
 primaryClass = {astro-ph.GA},
       adsurl = {https://ui.adsabs.harvard.edu/abs/2009Natur.462..480L}
}

@ARTICLE{lee2015,
       author = {{Lee}, Jae-Woo},
        title = "{Multiple Stellar Populations of Globular Clusters from Homogeneous Ca by Photometry. I. M22 (NGC 6656)}",
      journal = {\apjs},
         year = 2015,
        month = jul,
       volume = {219},
       number = {1},
          eid = {7},
        pages = {7},
          doi = {10.1088/0067-0049/219/1/7},
archivePrefix = {arXiv},
       eprint = {1506.00116},
 primaryClass = {astro-ph.GA},
       adsurl = {https://ui.adsabs.harvard.edu/abs/2015ApJS..219....7L}
}

@ARTICLE{lee2020,
       author = {{Lee}, Jae-Woo},
        title = "{Five Stellar Populations in M22 (NGC 6656)}",
      journal = {\apjl},
         year = 2020,
        month = jan,
       volume = {888},
       number = {1},
          eid = {L6},
        pages = {L6},
          doi = {10.3847/2041-8213/ab60b2},
archivePrefix = {arXiv},
       eprint = {2001.00679},
 primaryClass = {astro-ph.GA},
       adsurl = {https://ui.adsabs.harvard.edu/abs/2020ApJ...888L...6L}
}

@ARTICLE{lee2023,
       author = {{Lee}, Jae-Woo},
        title = "{Carbon Abundance of Globular Cluster M22 (NGC 6656) and the Surface Carbon Depletion Rates of the Milky Way Globular Clusters}",
      journal = {\apjl},
         year = 2023,
        month = jun,
       volume = {950},
       number = {1},
          eid = {L6},
        pages = {L6},
          doi = {10.3847/2041-8213/acd76b},
archivePrefix = {arXiv},
       eprint = {2306.04391},
 primaryClass = {astro-ph.GA},
       adsurl = {https://ui.adsabs.harvard.edu/abs/2023ApJ...950L...6L}
}

@ARTICLE{lehnert1991,
       author = {{Lehnert}, M.~D. and {Bell}, R.~A. and {Cohen}, J.~G.},
        title = "{Abundances in the Red Giants of M13 and M22}",
      journal = {\apj},
         year = 1991,
        month = feb,
       volume = {367},
        pages = {514},
          doi = {10.1086/169648},
       adsurl = {https://ui.adsabs.harvard.edu/abs/1991ApJ...367..514L}
}

@ARTICLE{lindegren2018,
       author = {{Lindegren}, L. and {Hern{\'a}ndez}, J. and {Bombrun}, A. and {Klioner}, S. and {Bastian}, U. and {Ramos-Lerate}, M. and {de Torres}, A. and {Steidelm{\"u}ller}, H. and {Stephenson}, C. and {Hobbs}, D. and {Lammers}, U. and {Biermann}, M. and {Geyer}, R. and {Hilger}, T. and {Michalik}, D. and {Stampa}, U. and {McMillan}, P.~J. and {Casta{\~n}eda}, J. and {Clotet}, M. and {Comoretto}, G. and {Davidson}, M. and {Fabricius}, C. and {Gracia}, G. and {Hambly}, N.~C. and {Hutton}, A. and {Mora}, A. and {Portell}, J. and {van Leeuwen}, F. and {Abbas}, U. and {Abreu}, A. and {Altmann}, M. and {Andrei}, A. and {Anglada}, E. and {Balaguer-N{\'u}{\~n}ez}, L. and {Barache}, C. and {Becciani}, U. and {Bertone}, S. and {Bianchi}, L. and {Bouquillon}, S. and {Bourda}, G. and {Br{\"u}semeister}, T. and {Bucciarelli}, B. and {Busonero}, D. and {Buzzi}, R. and {Cancelliere}, R. and {Carlucci}, T. and {Charlot}, P. and {Cheek}, N. and {Crosta}, M. and {Crowley}, C. and {de Bruijne}, J. and {de Felice}, F. and {Drimmel}, R. and {Esquej}, P. and {Fienga}, A. and {Fraile}, E. and {Gai}, M. and {Garralda}, N. and {Gonz{\'a}lez-Vidal}, J.~J. and {Guerra}, R. and {Hauser}, M. and {Hofmann}, W. and {Holl}, B. and {Jordan}, S. and {Lattanzi}, M.~G. and {Lenhardt}, H. and {Liao}, S. and {Licata}, E. and {Lister}, T. and {L{\"o}ffler}, W. and {Marchant}, J. and {Martin-Fleitas}, J. -M. and {Messineo}, R. and {Mignard}, F. and {Morbidelli}, R. and {Poggio}, E. and {Riva}, A. and {Rowell}, N. and {Salguero}, E. and {Sarasso}, M. and {Sciacca}, E. and {Siddiqui}, H. and {Smart}, R.~L. and {Spagna}, A. and {Steele}, I. and {Taris}, F. and {Torra}, J. and {van Elteren}, A. and {van Reeven}, W. and {Vecchiato}, A.},
        title = "{Gaia Data Release 2. The astrometric solution}",
      journal = {\aap},
         year = 2018,
        month = aug,
       volume = {616},
          eid = {A2},
        pages = {A2},
          doi = {10.1051/0004-6361/201832727},
archivePrefix = {arXiv},
       eprint = {1804.09366},
 primaryClass = {astro-ph.IM},
       adsurl = {https://ui.adsabs.harvard.edu/abs/2018A&A...616A...2L}
}

@ARTICLE{marino2009,
       author = {{Marino}, A.~F. and {Milone}, A.~P. and {Piotto}, G. and {Villanova}, S. and {Bedin}, L.~R. and {Bellini}, A. and {Renzini}, A.},
        title = "{A double stellar generation in the globular cluster NGC 6656 (M 22). Two stellar groups with different iron and s-process element abundances}",
      journal = {\aap},
         year = 2009,
        month = oct,
       volume = {505},
       number = {3},
        pages = {1099-1113},
          doi = {10.1051/0004-6361/200911827},
archivePrefix = {arXiv},
       eprint = {0905.4058},
 primaryClass = {astro-ph.SR},
       adsurl = {https://ui.adsabs.harvard.edu/abs/2009A&A...505.1099M}
}

@ARTICLE{marino2011a,
       author = {{Marino}, A.~F. and {Milone}, A.~P. and {Piotto}, G. and {Villanova}, S. and {Gratton}, R. and {D'Antona}, F. and {Anderson}, J. and {Bedin}, L.~R. and {Bellini}, A. and {Cassisi}, S. and {Geisler}, D. and {Renzini}, A. and {Zoccali}, M.},
        title = "{Sodium-Oxygen Anticorrelation and Neutron-capture Elements in Omega Centauri Stellar Populations}",
      journal = {\apj},
         year = 2011,
        month = apr,
       volume = {731},
       number = {1},
          eid = {64},
        pages = {64},
          doi = {10.1088/0004-637X/731/1/64},
archivePrefix = {arXiv},
       eprint = {1102.1653},
 primaryClass = {astro-ph.SR},
       adsurl = {https://ui.adsabs.harvard.edu/abs/2011ApJ...731...64M}
}

@ARTICLE{marino2011b,
       author = {{Marino}, A.~F. and {Sneden}, C. and {Kraft}, R.~P. and {Wallerstein}, G. and {Norris}, J.~E. and {Da Costa}, G. and {Milone}, A.~P. and {Ivans}, I.~I. and {Gonzalez}, G. and {Fulbright}, J.~P. and {Hilker}, M. and {Piotto}, G. and {Zoccali}, M. and {Stetson}, P.~B.},
        title = "{The two metallicity groups of the globular cluster M 22: a chemical perspective}",
      journal = {\aap},
         year = 2011,
        month = aug,
       volume = {532},
          eid = {A8},
        pages = {A8},
          doi = {10.1051/0004-6361/201116546},
archivePrefix = {arXiv},
       eprint = {1105.1523},
 primaryClass = {astro-ph.SR},
       adsurl = {https://ui.adsabs.harvard.edu/abs/2011A&A...532A...8M}
}

@ARTICLE{marino2011c,
       author = {{Marino}, A.~F. and {Villanova}, S. and {Milone}, A.~P. and {Piotto}, G. and {Lind}, K. and {Geisler}, D. and {Stetson}, P.~B.},
        title = "{Sodium-Oxygen Anticorrelation Among Horizontal Branch Stars in the Globular Cluster M4}",
      journal = {\apjl},
         year = 2011,
        month = apr,
       volume = {730},
       number = {2},
          eid = {L16},
        pages = {L16},
          doi = {10.1088/2041-8205/730/2/L16},
archivePrefix = {arXiv},
       eprint = {1012.4931},
 primaryClass = {astro-ph.SR},
       adsurl = {https://ui.adsabs.harvard.edu/abs/2011ApJ...730L..16M}
}

@ARTICLE{marino2013,
       author = {{Marino}, A.~F. and {Milone}, A.~P. and {Lind}, K.},
        title = "{Horizontal Branch Morphology and Multiple Stellar Populations in the Anomalous Globular Cluster M 22}",
      journal = {\apj},
         year = 2013,
        month = may,
       volume = {768},
       number = {1},
          eid = {27},
        pages = {27},
          doi = {10.1088/0004-637X/768/1/27},
archivePrefix = {arXiv},
       eprint = {1302.5870},
 primaryClass = {astro-ph.SR},
       adsurl = {https://ui.adsabs.harvard.edu/abs/2013ApJ...768...27M}
}

@ARTICLE{marino2015,
       author = {{Marino}, A.~F. and {Milone}, A.~P. and {Karakas}, A.~I. and {Casagrande}, L. and {Yong}, D. and {Shingles}, L. and {Da Costa}, G. and {Norris}, J.~E. and {Stetson}, P.~B. and {Lind}, K. and {Asplund}, M. and {Collet}, R. and {Jerjen}, H. and {Sbordone}, L. and {Aparicio}, A. and {Cassisi}, S.},
        title = "{Iron and s-elements abundance variations in NGC 5286: comparison with `anomalous' globular clusters and Milky Way satellites}",
      journal = {\mnras},
         year = 2015,
        month = jun,
       volume = {450},
       number = {1},
        pages = {815-845},
          doi = {10.1093/mnras/stv420},
archivePrefix = {arXiv},
       eprint = {1502.07438},
 primaryClass = {astro-ph.SR},
       adsurl = {https://ui.adsabs.harvard.edu/abs/2015MNRAS.450..815M}
}

@ARTICLE{marino2019,
       author = {{Marino}, A.~F. and {Milone}, A.~P. and {Renzini}, A. and {D'Antona}, F. and {Anderson}, J. and {Bedin}, L.~R. and {Bellini}, A. and {Cordoni}, G. and {Lagioia}, E.~P. and {Piotto}, G. and {Tailo}, M.},
        title = "{The Hubble Space Telescope UV Legacy Survey of Galactic Globular Clusters - XIX. A chemical tagging of the multiple stellar populations over the chromosome maps}",
      journal = {\mnras},
         year = 2019,
        month = aug,
       volume = {487},
       number = {3},
        pages = {3815-3844},
          doi = {10.1093/mnras/stz1415},
archivePrefix = {arXiv},
       eprint = {1904.05180},
 primaryClass = {astro-ph.SR},
       adsurl = {https://ui.adsabs.harvard.edu/abs/2019MNRAS.487.3815M}
}

@ARTICLE{mason2025,
       author = {{Mason}, Andrew C. and {Schiavon}, Ricardo P. and {Kamann}, Sebastian and {Smith}, Verne V. and {Horta}, Danny and {Anguiano}, Borja and {Cunha}, Katia and {M{\'e}sz{\'a}ros}, Szabolcs and {Majewski}, Steven R. and {O'Connell}, Robert W. and {Allende Prieto}, Carlos and {Saracino}, Sara},
        title = "{Chemical tagging with APOGEE, Gaia, MUSE, and HST: constraints on the formation of {\ensuremath{\omega}} Centauri}",
      journal = {\mnras},
         year = 2026,
        month = apr,
       volume = {547},
       number = {4},
          eid = {stag433},
        pages = {stag433},
          doi = {10.1093/mnras/stag433},
archivePrefix = {arXiv},
       eprint = {2504.06341},
 primaryClass = {astro-ph.GA},
       adsurl = {https://ui.adsabs.harvard.edu/abs/2026MNRAS.547ag433M}
}

@ARTICLE{matteucci2021,
       author = {{Matteucci}, Francesca},
        title = "{Modelling the chemical evolution of the Milky Way}",
      journal = {\aapr},
         year = 2021,
        month = dec,
       volume = {29},
       number = {1},
          eid = {5},
        pages = {5},
          doi = {10.1007/s00159-021-00133-8},
archivePrefix = {arXiv},
       eprint = {2106.13145},
 primaryClass = {astro-ph.GA},
       adsurl = {https://ui.adsabs.harvard.edu/abs/2021A&ARv..29....5M}
}

@ARTICLE{mcfarland2013,
       author = {{McFarland}, John P. and {Verdoes-Kleijn}, Gijs and {Sikkema}, Gert and {Helmich}, Ewout M. and {Boxhoorn}, Danny R. and {Valentijn}, Edwin A.},
        title = "{The Astro-WISE optical image pipeline. Development and implementation}",
      journal = {Experimental Astronomy},
         year = 2013,
        month = jan,
       volume = {35},
       number = {1-2},
        pages = {45-78},
          doi = {10.1007/s10686-011-9266-x},
       adsurl = {https://ui.adsabs.harvard.edu/abs/2013ExA....35...45M}
}

@ARTICLE{mckenzie2022,
       author = {{McKenzie}, M. and {Yong}, D. and {Marino}, A.~F. and {Monty}, S. and {Wang}, E. and {Karakas}, A.~I. and {Milone}, A.~P. and {Legnardi}, M.~V. and {Roederer}, I.~U. and {Martell}, S. and {Horta}, D.},
        title = "{The complex stellar system M 22: confirming abundance variations with high precision differential measurements}",
      journal = {\mnras},
         year = 2022,
        month = nov,
       volume = {516},
       number = {3},
        pages = {3515-3531},
          doi = {10.1093/mnras/stac2254},
       adsurl = {https://ui.adsabs.harvard.edu/abs/2022MNRAS.516.3515M}
}

@ARTICLE{mckenzie2024,
       author = {{McKenzie}, M. and {Yong}, D. and {Karakas}, A.~I. and {Wang}, E. and {Monty}, S. and {Marino}, A.~F. and {Milone}, A.~P. and {Nordlander}, T. and {Mura-Guzm{\'a}n}, A. and {Martell}, S. and {Carlos}, M.},
        title = "{The complex stellar system M 22: constraining the chemical enrichment from AGB stars using magnesium isotope ratios}",
      journal = {\mnras},
         year = 2024,
        month = jan,
       volume = {527},
       number = {3},
        pages = {7940-7955},
          doi = {10.1093/mnras/stad2999},
       adsurl = {https://ui.adsabs.harvard.edu/abs/2024MNRAS.527.7940M}
}

@ARTICLE{milone2008,
       author = {{Milone}, A.~P. and {Bedin}, L.~R. and {Piotto}, G. and {Anderson}, J. and {King}, I.~R. and {Sarajedini}, A. and {Dotter}, A. and {Chaboyer}, B. and {Mar{\'\i}n-Franch}, A. and {Majewski}, S. and {Aparicio}, A. and {Hempel}, M. and {Paust}, N.~E.~Q. and {Reid}, I.~N. and {Rosenberg}, A. and {Siegel}, M.},
        title = "{The ACS Survey of Galactic Globular Clusters. III. The Double Subgiant Branch of NGC 1851}",
      journal = {\apj},
         year = 2008,
        month = jan,
       volume = {673},
       number = {1},
        pages = {241-250},
          doi = {10.1086/524188},
archivePrefix = {arXiv},
       eprint = {0709.3762},
 primaryClass = {astro-ph},
       adsurl = {https://ui.adsabs.harvard.edu/abs/2008ApJ...673..241M}
}

@ARTICLE{milone2012b,
       author = {{Milone}, A.~P. and {Piotto}, G. and {Bedin}, L.~R. and {Aparicio}, A. and {Anderson}, J. and {Sarajedini}, A. and {Marino}, A.~F. and {Moretti}, A. and {Davies}, M.~B. and {Chaboyer}, B. and {Dotter}, A. and {Hempel}, M. and {Mar{\'\i}n-Franch}, A. and {Majewski}, S. and {Paust}, N.~E.~Q. and {Reid}, I.~N. and {Rosenberg}, A. and {Siegel}, M.},
        title = "{The ACS survey of Galactic globular clusters. XII. Photometric binaries along the main sequence}",
      journal = {\aap},
         year = 2012,
        month = apr,
       volume = {540},
          eid = {A16},
        pages = {A16},
          doi = {10.1051/0004-6361/201016384},
archivePrefix = {arXiv},
       eprint = {1111.0552},
 primaryClass = {astro-ph.SR},
       adsurl = {https://ui.adsabs.harvard.edu/abs/2012A&A...540A..16M}
}

@ARTICLE{milone2017,
       author = {{Milone}, A.~P. and {Piotto}, G. and {Renzini}, A. and {Marino}, A.~F. and {Bedin}, L.~R. and {Vesperini}, E. and {D'Antona}, F. and {Nardiello}, D. and {Anderson}, J. and {King}, I.~R. and {Yong}, D. and {Bellini}, A. and {Aparicio}, A. and {Barbuy}, B. and {Brown}, T.~M. and {Cassisi}, S. and {Ortolani}, S. and {Salaris}, M. and {Sarajedini}, A. and {van der Marel}, R.~P.},
        title = "{The Hubble Space Telescope UV Legacy Survey of Galactic globular clusters - IX. The Atlas of multiple stellar populations}",
      journal = {\mnras},
         year = 2017,
        month = jan,
       volume = {464},
       number = {3},
        pages = {3636-3656},
          doi = {10.1093/mnras/stw2531},
archivePrefix = {arXiv},
       eprint = {1610.00451},
 primaryClass = {astro-ph.SR},
       adsurl = {https://ui.adsabs.harvard.edu/abs/2017MNRAS.464.3636M}
}

@ARTICLE{milone2018,
       author = {{Milone}, A.~P. and {Marino}, A.~F. and {Renzini}, A. and {D'Antona}, F. and {Anderson}, J. and {Barbuy}, B. and {Bedin}, L.~R. and {Bellini}, A. and {Brown}, T.~M. and {Cassisi}, S. and {Cordoni}, G. and {Lagioia}, E.~P. and {Nardiello}, D. and {Ortolani}, S. and {Piotto}, G. and {Sarajedini}, A. and {Tailo}, M. and {van der Marel}, R.~P. and {Vesperini}, E.},
        title = "{The Hubble Space Telescope UV legacy survey of galactic globular clusters - XVI. The helium abundance of multiple populations}",
      journal = {\mnras},
         year = 2018,
        month = dec,
       volume = {481},
       number = {4},
        pages = {5098-5122},
          doi = {10.1093/mnras/sty2573},
archivePrefix = {arXiv},
       eprint = {1809.05006},
 primaryClass = {astro-ph.SR},
       adsurl = {https://ui.adsabs.harvard.edu/abs/2018MNRAS.481.5098M}
}

@ARTICLE{milone2022,
       author = {{Milone}, Antonino P. and {Marino}, Anna F.},
        title = "{Multiple Populations in Star Clusters}",
      journal = {Universe},
         year = 2022,
        month = jun,
       volume = {8},
       number = {7},
          eid = {359},
        pages = {359},
          doi = {10.3390/universe8070359},
archivePrefix = {arXiv},
       eprint = {2206.10564},
 primaryClass = {astro-ph.GA},
       adsurl = {https://ui.adsabs.harvard.edu/abs/2022Univ....8..359M}
}

@ARTICLE{norris1983,
       author = {{Norris}, J. and {Freeman}, K.~C.},
        title = "{The chemical inhomogeneity of M 22.}",
      journal = {\apj},
         year = 1983,
        month = mar,
       volume = {266},
        pages = {130-143},
          doi = {10.1086/160764},
       adsurl = {https://ui.adsabs.harvard.edu/abs/1983ApJ...266..130N}
}

@ARTICLE{olszewki2009,
       author = {{Olszewski}, Edward W. and {Saha}, Abhijit and {Knezek}, Patricia and {Subramaniam}, Annapurni and {de Boer}, Thomas and {Seitzer}, Patrick},
        title = "{A 500 Parsec Halo Surrounding the Galactic Globular NGC 1851}",
      journal = {\aj},
         year = 2009,
        month = dec,
       volume = {138},
       number = {6},
        pages = {1570-1576},
          doi = {10.1088/0004-6256/138/6/1570},
archivePrefix = {arXiv},
       eprint = {0909.1755},
 primaryClass = {astro-ph.GA},
       adsurl = {https://ui.adsabs.harvard.edu/abs/2009AJ....138.1570O}
}

@ARTICLE{pedregosa2011,
       author = {{Pedregosa}, Fabian and {Varoquaux}, Ga{\"e}l and {Gramfort}, Alexandre and {Michel}, Vincent and {Thirion}, Bertrand and {Grisel}, Olivier and {Blondel}, Mathieu and {M{\"u}ller}, Andreas and {Nothman}, Joel and {Louppe}, Gilles and {Prettenhofer}, Peter and {Weiss}, Ron and {Dubourg}, Vincent and {Vanderplas}, Jake and {Passos}, Alexandre and {Cournapeau}, David and {Brucher}, Matthieu and {Perrot}, Matthieu and {Duchesnay}, {\'E}douard},
        title = "{Scikit-learn: Machine Learning in Python}",
      journal = {Journal of Machine Learning Research},
         year = 2011,
        month = oct,
       volume = {12},
        pages = {2825-2830},
          doi = {10.48550/arXiv.1201.0490},
archivePrefix = {arXiv},
       eprint = {1201.0490},
 primaryClass = {cs.LG},
       adsurl = {https://ui.adsabs.harvard.edu/abs/2011JMLR...12.2825P}
}

@ARTICLE{piotto2012,
       author = {{Piotto}, G. and {Milone}, A.~P. and {Anderson}, J. and {Bedin}, L.~R. and {Bellini}, A. and {Cassisi}, S. and {Marino}, A.~F. and {Aparicio}, A. and {Nascimbeni}, V.},
        title = "{Hubble Space Telescope Reveals Multiple Sub-giant Branch in Eight Globular Clusters}",
      journal = {\apj},
         year = 2012,
        month = nov,
       volume = {760},
       number = {1},
          eid = {39},
        pages = {39},
          doi = {10.1088/0004-637X/760/1/39},
archivePrefix = {arXiv},
       eprint = {1208.1873},
 primaryClass = {astro-ph.SR},
       adsurl = {https://ui.adsabs.harvard.edu/abs/2012ApJ...760...39P}
}

@ARTICLE{renzini2023,
       author = {{Renzini}, Alvio},
        title = "{A transient overcooling in the early Universe? Clues from globular clusters formation}",
      journal = {\mnras},
         year = 2023,
        month = oct,
       volume = {525},
       number = {1},
        pages = {L117-L120},
          doi = {10.1093/mnrasl/slad091},
archivePrefix = {arXiv},
       eprint = {2305.14476},
 primaryClass = {astro-ph.GA},
       adsurl = {https://ui.adsabs.harvard.edu/abs/2023MNRAS.525L.117R}
}

@ARTICLE{shetrone2026,
       author = {{Shetrone}, Matthew and {Beaton}, Rachael L. and {Hayes}, Christian R. and {Hasselquist}, Sten and {Simon}, Joshua D. and {Holtzman}, Jon A. and {Cunha}, Katia and {Majewski}, Steven R. and {Sobeck}, Jennifer and {Schiavon}, Ricardo and {Masseron}, Thomas and {Smith}, Verne V. and {Nidever}, David L.},
        title = "{The Apache Point Observatory Extra-galactic Evolution Experiment (APOeGEE): Chemical Abundance Trends for Seven Dwarf Spheroidal Galaxies in the APOGEE Survey}",
      journal = {\apj},
         year = 2026,
        month = feb,
       volume = {998},
       number = {1},
          eid = {117},
        pages = {117},
          doi = {10.3847/1538-4357/ae27c5},
archivePrefix = {arXiv},
       eprint = {2511.04365},
 primaryClass = {astro-ph.GA},
       adsurl = {https://ui.adsabs.harvard.edu/abs/2026ApJ...998..117S}
}

@ARTICLE{souza2026,
       author = {{Souza}, Stefano and {Neumayer}, Nadine and {Seth}, Anil C. and {Wang}, Zixian and {Clontz}, Callie and {H{\"a}berle}, Maximilian and {Nitschai}, Maria S. and {Smith}, Peter J. and {Matsuno}, Tadafumi and {Guiglion}, Guillaume and {Feldmeier-Krause}, Anja and {Kacharov}, Nikolay and {van de Ven}, Glenn and {Li}, Jiadong and {Libralato}, Mattia and {Bellini}, Andrea and {Milone}, Antonino P. and {Alfaro-Cuello}, Mayte},
        title = "{oMEGACat. X. Shedding light on the disrupted dwarf galaxy of Omega Centauri}",
      journal = {arXiv e-prints},
         year = 2026,
        month = mar,
          eid = {arXiv:2603.23589},
        pages = {arXiv:2603.23589},
          doi = {10.48550/arXiv.2603.23589},
archivePrefix = {arXiv},
       eprint = {2603.23589},
 primaryClass = {astro-ph.GA},
       adsurl = {https://ui.adsabs.harvard.edu/abs/2026arXiv260323589S}
}

@ARTICLE{stetson2019,
       author = {{Stetson}, P.~B. and {Pancino}, E. and {Zocchi}, A. and {Sanna}, N. and {Monelli}, M.},
        title = "{Homogeneous photometry - VII. Globular clusters in the Gaia era}",
      journal = {\mnras},
         year = 2019,
        month = may,
       volume = {485},
       number = {3},
        pages = {3042-3063},
          doi = {10.1093/mnras/stz585},
archivePrefix = {arXiv},
       eprint = {1902.09925},
 primaryClass = {astro-ph.SR},
       adsurl = {https://ui.adsabs.harvard.edu/abs/2019MNRAS.485.3042S}
}

@ARTICLE{tailo2020,
       author = {{Tailo}, M. and {Milone}, A.~P. and {Lagioia}, E.~P. and {D'Antona}, F. and {Marino}, A.~F. and {Vesperini}, E. and {Caloi}, V. and {Ventura}, P. and {Dondoglio}, E. and {Cordoni}, G.},
        title = "{Mass-loss along the red giant branch in 46 globular clusters and their multiple populations}",
      journal = {\mnras},
         year = 2020,
        month = nov,
       volume = {498},
       number = {4},
        pages = {5745-5771},
          doi = {10.1093/mnras/staa2639},
archivePrefix = {arXiv},
       eprint = {2009.01080},
 primaryClass = {astro-ph.SR},
       adsurl = {https://ui.adsabs.harvard.edu/abs/2020MNRAS.498.5745T}
}

@ARTICLE{ventura2022,
       author = {{Ventura}, Paolo and {Dell'Agli}, Flavia and {Tailo}, Marco and {Castellani}, Marco and {Marini}, Ester and {Tosi}, Silvia and {Di Criscienzo}, Marcella},
        title = "{Nucleosynthesis, Mixing Processes, and Gas Pollution from AGB Stars}",
      journal = {Universe},
         year = 2022,
        month = jan,
       volume = {8},
       number = {1},
          eid = {45},
        pages = {45},
          doi = {10.3390/universe8010045},
       adsurl = {https://ui.adsabs.harvard.edu/abs/2022Univ....8...45V}
}

@ARTICLE{yong2008,
       author = {{Yong}, David and {Grundahl}, Frank},
        title = "{An Abundance Analysis of Bright Giants in the Globular Cluster NGC 1851}",
      journal = {\apjl},
         year = 2008,
        month = jan,
       volume = {672},
       number = {1},
        pages = {L29},
          doi = {10.1086/525850},
archivePrefix = {arXiv},
       eprint = {0711.1394},
 primaryClass = {astro-ph},
       adsurl = {https://ui.adsabs.harvard.edu/abs/2008ApJ...672L..29Y}
}

@ARTICLE{xu2026,
       author = {{Xu}, Cheng and {Qiao}, Yi and {Tang}, Baitian and {Fern{\'a}ndez-Trincado}, Jos{\'e} G. and {Yan}, Zhiqiang and {Huang}, Ruoyun and {Geisler}, Doug},
        title = "{APOGEE chemical abundances of stars in the Milky Way satellites Fornax, Sextans, Draco, and Carina}",
      journal = {\aap},
         year = 2026,
        month = apr,
       volume = {708},
          eid = {A259},
        pages = {A259},
          doi = {10.1051/0004-6361/202555822},
archivePrefix = {arXiv},
       eprint = {2511.06820},
 primaryClass = {astro-ph.GA},
       adsurl = {https://ui.adsabs.harvard.edu/abs/2026A&A...708A.259X}
}

@ARTICLE{Asad24,
       author = {{Asa'd}, Randa and {Hernandez}, S. and {John}, J.~M. and {Alfaro-Cuello}, M. and {Wang}, Z. and {As'ad}, A. and {Vasini}, A. and {Matteucci}, F.},
        title = "{NGC 1856: Using Machine Learning Techniques to Uncover Detailed Stellar Abundances from MUSE Data}",
      journal = {\aj},
         year = 2024,
        month = jun,
       volume = {167},
       number = {6},
          eid = {265},
        pages = {265},
          doi = {10.3847/1538-3881/ad3f1b},
archivePrefix = {arXiv},
       eprint = {2404.15527},
 primaryClass = {astro-ph.GA},
       adsurl = {https://ui.adsabs.harvard.edu/abs/2024AJ....167..265A}
}
\bibliographystyle{aasjournalv7.1}



\end{document}